%% file: sample631.tex
\documentclass[twocolumn]{aastex631}
\pdfoutput=1

\submitjournal{ApJ}
\shorttitle{Escape Mass of AS1063}
\shortauthors{Rodriguez}
\graphicspath{{./}{figures/}}
\usepackage{natbib}
\usepackage [autostyle, english = american]{csquotes}
\usepackage{enumerate}
\usepackage{mathtools}
\usepackage{amsmath}
\usepackage{amssymb}
\usepackage{appendix}
\usepackage{soul}
\usepackage{float}
\usepackage [english]{babel}
\usepackage [autostyle, english = american]{csquotes}
\usepackage{enumerate}
\usepackage{hyperref}
\def\arnote#1{{\color{black}{#1}}}

\usepackage{threeparttable}
\renewcommand{\tnote}[1]{\textsuperscript{\textbf{#1}}}

\begin{document}
\title{Escape Velocity Mass of Abell S1063}
\email{alexcrod@umich.edu}

\author[0000-0002-8813-9116]{Alexander Rodriguez}
\affiliation{University of Michigan, Ann Arbor \\
Department of Astronomy}
\author[0000-0002-1365-0841]{Christopher J. Miller}
\affiliation{University of Michigan, Ann Arbor \\
Department of Astronomy}
\affiliation{University of Michigan, Ann Arbor \\
Department of Physics}
\author[0000-0001-7483-0945]{Vitali Halenka}
\affiliation{Independent Researcher}
\author[0000-0001-6356-7424]{Anthony Kremin}
\affiliation{Lawrence Berkeley National Laboratory, Berkeley, CA 94720}

\begin{abstract}
{We measure the radius-velocity phase-space edge profile for Abell S1063 using galaxy redshifts from \citet{karman+2015} and \citet{mercurio+2021}. Combined with a cosmological model and after accounting for interlopers and sampling effects, we infer the escape velocity profile. Using the Poisson equation, we then directly constrain the gravitational potential profile and find excellent agreement between three different density models. For the NFW profile, we find log$_{10}$(M$_{200},{\rm crit}$)=
$15.40^{+0.06}_{-0.12}$M$_{\odot}$, consistent to within $1\sigma$ of six recently published lensing masses. We argue that this consistency is due to the fact that the escape technique shares no common systematics with lensing other than radial binning. These masses are 2-4$\sigma$ lower than estimates using X-ray data, in addition to the \citet{gomez+2012} velocity dispersion estimate. We measure the 1D velocity dispersion within r$_{200}$ to be $\sigma_{v} = 1477^{+87}_{-99}$ km/s, 
which combined with our escape velocity mass, brings the dispersion for AS1063 in-line with hydrodynamic cosmological simulations for the first time.} 

\end{abstract}
\keywords{Galaxy clusters, Dark energy, Cosmology, Gravitational lensing}

\section{Introduction}
\label{sec:Introduction}
Galaxy clusters represent the most massive gravitationally bound structures having formed to date. Outside of their cores, these systems are unique in that their evolution is determined only through the combined effects of gravity and the expanding spacetime from the accelerated universal expansion on $\sim\text{Mpc}$ scales.

In the current $\Lambda$CDM framework, the cluster potential is dominated by dark matter (hereafter DM). A galaxy in the cluster feels an inward force from not only the DM, gas, and galaxies, but also an outward force due to the accelerated universal expansion. In a non-accelerating universe, \arnote{dynamical tracers of the cluster potential can escape a cluster if it can somehow reach a speed above the escape velocity}, given by
\begin{align}
\label{eq:vesc}
    v_{\text{esc}}\arnote{(r)}&=\sqrt{-2\Psi(r)},
\end{align}
where $\Psi(r)$ is the potential from matter alone. 

\arnote{In an accelerating universe}, \citet{nandra+2012} showed that there is a radius ($r_\text{{eq}}$) at which the \arnote{inward} force on a test particle by a massive object is perfectly balanced by the effective \arnote{outward} force of the accelerating space time. \arnote{In the accelerating case,} \citet{Behroozi+2013} showed that a form of Equation \ref{eq:vesc} holds as well, where the matter-only potential $\Psi(r)$ is replaced by the effective potential, $\Phi(r)$, which includes contributions from both gravity and the accelerated expansion. \arnote{The key difference being that the escape speed relative to the clusters becomes zero at $r_\text{{eq}}$.}


Arguably, the most important factor in determining the evolution of a cluster is its mass. \citet{diaferio+1997} showed that it is possible to relate the mass of a cluster to its escape velocity through construction of a radius/velocity phase-space diagram. Within the phase-space, particles, sub-halos, and galaxies are confined to regions within which they are bound, confined by the phase-space edge, a \arnote{``caustic''}, or sharp cut-off in the phase-space with a velocity at a corresponding radius given by Equation \ref{eq:vesc}. 


\arnote{Simulations have been used to show that} the escape velocity can be measured from a clearly defined edge in the radius/velocity phase-space diagram. \arnote{Except for a very small number of escaping tracers,} only those with the maximum possible radial or tangential 1D speed will contribute to this edge \citep{Behroozi+2013}. The power of utilizing the observed $v_\text{esc}(r)$ is in its direct connection to the total potential, enabling cluster mass estimations, tests of gravity on the largest scales in the weak field limit, and placing constraints on the $\Lambda$CDM cosmological parameters \citep{miller+2016, stark2016, stark2017}. 
\arnote{Notably, the cosmic acceleration results in a smaller escape velocity for a given mass than one would expect in a universe without acceleration \citep{diaferio+1997, Behroozi+2013, miller+2016, stark2017}. }

\arnote{Prior work associated with utilizing observed phase-spaces for measuring cluster masses has required cosmological simulations to calibrate ``caustic'' surfaces \citep{diaferio+1997, Diaferio1999, Diaferio2005, Serra2011, Geller13, Gifford_2013, Pizzardo2023a, Pizzardo2023b}. These caustics are then related to cluster masses or mass profiles. However, \citet{Gifford_2017} and \citet{halenka+2020} employed direct measurements of the observed maximum phase-space peculiar velocity edge. \citet{Gifford_2017} defined the edge as the 90$^{th}$ percentile of the peculiar velocity distribution to infer average cluster masses. \citet{halenka+2020} measured the absolute maximum velocity in projected data to infer the 3D escape profile. We will use Abell S1063 as an observational test of how well the observed phase-space maximum edge profile relates to the cluster mass.}

Abell S1063 (hereafter AS1063)--also referred to as RXC J2248.7-4431--- is a massive, luminous, cluster, located at a redshift of $z=0.3475$ \citep{abell1989,Bohringer+2004}, and is part of the Cluster Lensing And Supernova survey with Hubble (HST CLASH) program \citep{Postman+2012,lotz+2017}. \arnote{This cluster has had its mass measured many times using independent techniques, making it an ideal choice for our analysis.}

\citet{gomez+2012} performed an in-depth investigation of properties of AS1063 by measuring its X-ray temperature and velocity dispersion (given in Table \ref{table:Summaray_mass_estimates}). These were obtained through Gemini observations, and found a corresponding mass of $\log{M_{200c}} \sim 15.6 \log(M_{\odot})$ \footnote{$M_{200c}$ refers to the mass within which the mean density of the cluster is $200\times\rho_c(z)$, where $\rho_c(z)$ is the critical density of the universe at redshift $z$.}, extremely high relative to most known clusters. 

\citet{gruen2013} subsequently performed an analysis using weak lensing, and found a mass nearly $\sim2\times$ lower (again see Table \ref{table:Summaray_mass_estimates} for further details), attributing the discrepancy to bulk motion of different galaxy populations along the line-of-sight, \arnote{evidence for a} dynamically non-relaxed system. 

\citet{Umetsu2014,Umetsu_2016} repeated the analysis of \citet{gruen2013} with the same CLASH-VLT data, but with independent shear calibration measurements and photometry, resulting in findings which more closely matched \citet{gruen2013} than \citet{gomez+2012}. 

\citet{sartoris+2020} then combined the CLASH data with \citet{caminha+2016}, where 16 additional background sources for the cluster had been discovered. These data were used for a shear + magnification lensing analysis, consistent with the previous lensing estimates from \citet{gruen2013,Umetsu2014,Umetsu_2016}. 

Due to its extreme X-ray luminosity, \citet{comis+2011,gomez+2012} also studied the cluster using its X-ray properties, using data from Chandra. The resulting X-ray masses were similar to the dynamical mass using $\sigma_v$ from \citet{gomez+2012}, a systematic offset between the dynamical and X-ray masses compared to lensing estimates. \citet{Donahue14} repeated the X-ray analysis XMM-Newton and Chandra telescopes, and again found similar results for the X-ray mass.

Subsequently, the prominent SZ signal in the CMB sparked the Planck team \citep{planck+2011} to use AS1063 for SZ analyses. \citet{williamson+2011,Penna-Lima2017,Capasso2019} measured SZ masses which are significantly lower than the X-ray estimates and more consistent with the lensing results. 

Most recently, \citet{sartoris+2020} performed another dynamical analysis of AS1063, this time making use of MAMPOSSt \citep{Mamon+2013}. This technique involves inversion of the Jean's equation of dynamical equilibrium \citep{binney+2008} to perform a joint maximum likelihood fit to the velocity dispersion profile of the BCG and to the velocity distribution of cluster member galaxies. The resulting dynamical mass is higher than the lensing and SZ estimates, \arnote{although in better agreement than the dynamical estimate from \citet{gomez+2012} using the velocity dispersion.}

\citet{mercurio+2021} expanded upon these recent findings by combining CLASH-VLT observations with \citet{karman+2015}, and found \arnote{strong statistical} evidence of merging subclumps in the system. These subclumps were in addition to a bi-modal distribution of peculiar redshifts--further evidence that the cluster is not dynamically relaxed.

 Thanks to the CLASH-VLT spectroscopic program, the radius/velocity phase-space of AS1063 has been mapped with thousands of galaxies. Alongside the numerous prior independent mass measurements, these data enable us to conduct the first escape edge analysis of a cluster. Notably, we will not be modeling the internal structure of the phase-space data. In fact, only a few 10s of galaxies in the phase-space will ultimately contribute to the edge measurement and used in the statistical inference. We will use simulations and analytical phase-space representations of AS1063-like systems to quantify the accuracy and precision of our edge measurement.

The paper is structured as follows: in~\S\ref{sec:Theoretical Background}, we present a theoretical overview of the methodology of the escape edge framework. In~\S\ref{sec:Edge measurement}, we \arnote{perform a detailed measurement of the phase-space edge profile and its error. Mass inference based on the escape profile is presented in~\S\ref{subsec:mass_estimation}. In~\S\ref{sec:Velocity Dispersion Analysis},
we present a new estimate of the velocity dispersion where we model the many contributions to its uncertainty.} In~\S\ref{sec:discussion}, we discuss the astrophysical consequences of our findings, and implications.

Throughout this paper, we adopt $H_0 = 72 \, \text{km} \, \text{s}^{-1} \, \text{Mpc}^{-1}$, $\Omega_M=0.27$, $\Omega_{\Lambda}=0.73$. We call $r_{\Delta}$ the radius that encloses an average density $\Delta$ times the critical or mean density at the cluster redshift. In all plots and results, we use the critical threshold unless explicitly stated otherwise.

\begin{table}[t]
  \caption{Summary of the various mass estimation methods from previous work. The Intracluster Medium (ICM) masses require a hydrostatic equilibrium correction, which is incorporated in some of the ICM measurements below. Masses are measured with respect to $\rho_{\text{crit}}$. In Figure \ref{fig:v_los_data}, we show these mass estimates compared to our mass constraint from the escape velocity, in addition to the theoretical expectation from simulations.}
  \label{table:Summaray_mass_estimates}
  \centering
  \begin{tabular}{lcr}
 \hline\hline
AS1063 lensing log mass estimates & $M_{200c} (M_{\odot})$ \\  
\hline\hline
 Shear\tnote{1} \citep{gruen2013} & $15.37^{+0.11}_{-0.10}$ \\

 Shear \citep{Melchior2015} & $15.24^{+0.10}_{-0.10}$ \\
 
 Shear \citep{Klein2019} & $15.22^{+0.12}_{+0.10}$ \\

 + magnification\tnote{1} \citep{Umetsu2014} & $15.32^{+0.13}_{-0.18}$  \\
 + magnification \citep{sartoris+2020} & $15.34^{+0.10}_{-0.15}$  \\

 + mag + strong\tnote{1}  \citep{Umetsu_2016} & $15.28^{+0.13}_{-0.19}$  \\

\\
 \hline\hline
 AS1063 ICM HSE log mass estimates & $M_{200c} (M_{\odot})$ \\ \hline\hline

X-ray\tnote{2},\tnote{4}  (Chandra)
  \citep{Donahue14} & $15.60^{+0.04}_{-0.05}$  \\

X-ray\tnote{2},\tnote{4}  (XMM)
  \citep{Donahue14} & $15.52^{+0.02}_{-0.02}$  \\

 X-ray\tnote{3} (Chandra)
  \citep{sartoris+2020} & $15.57^{+0.07}_{-0.08}$  \\

 SZ\tnote{3},\tnote{5} (SPT)
\citep{williamson+2011}  & $15.34^{+0.08}_{-0.10}$  \\

 SZ\tnote{2},\tnote{6} (Planck)
\citep{Penna-Lima2017}  & $15.25^{+0.03}_{-0.01}$  \\

\\
 \hline\hline
AS1063 dynamical log mass estimates & $M_{200c} (M_{\odot})$ \\  
 \hline\hline

$\sigma_v$\tnote{7} 
  \citep{gomez+2012} & $15.60^{+0.15}_{-0.11}$  \\

 Jeans\tnote{8}
  \citep{sartoris+2020} & $15.46^{+0.04}_{-0.05}$  \\

 \\

  \end{tabular}

  \begin{tablenotes}[flushright, para]\footnotesize
  \item[1] We convert their $M_{200,c}$ in a $\Omega_M=0.27$ universe to a $M_{200,c}$ in a $\Omega_M=0.3$ universe.
  \item[2] This Work accounts for a hydrostatic bias correction.
  \item[3] This Work does not account for a hydrostatic bias correction.
  \item[4] We convert their $M_{2500,c}$ to $M_{200,c}$ using the \citep{Duffy_2008} mass-concentration relation.
  \item[5] From a Mass versus S/N scaling relation calibrated to simulations. We convert their $M_{200m}$ to w.r.t. to critical as described therein. We use the average of statistical and systematic errors. 
  \item[6] Calibrated using X-ray scaling relations. We convert their $M_{500,c}$ to $M_{200,c}$ using the \citep{Duffy_2008} mass-concentration relation.
  \item[7] They measure $\sigma_v$ to $\sim 750kpc$ and estimate what it should be at $r_{200}$ using N-body simulations. We use this ``de-biased core'' $\sigma_v$ in the \citet{evrard+2008} scaling relation to represent a dynamical mass.
  \item[8] This is the total mass: dark matter + gas.

  \end{tablenotes}
\end{table}

\section{Theoretical Background}
\label{sec:Theoretical Background}
\subsection{Escape dynamics}
\label{subsec: escape dynamics}
\citet{nandra+2012} studied the dynamical effects of an accelerating spacetime on a galaxy cluster in
the weak-field limit of general relativity. In this paradigm, we can relate the acceleration to the corresponding effective potential, $\Phi(r)$, with $a_{\text{eff}}=\nabla \Phi(r)$, where $a_{\text{eff}}$ is the effective acceleration on the particle
\begin{align}
\label{eq:nabla_phi}
    a_{\text{eff}}&=\nabla\Phi(r)=\nabla\Psi(r)+q(z)H^2(z) r\hat{r}
\end{align}
Here, $\Psi(r)$ is the potential due to dark matter, ICM, and galaxies, and $H(z)=H_0\sqrt{\Omega_\Lambda+\Omega_M(1+z)^3}$, with $q(z)=\frac{1}{2}\Omega_M(z)-\Omega_\Lambda(z)$ (assuming a flat universe, or $\Omega_k=0$) \citep{sandage+1970}. The second term in Equation \ref{eq:nabla_phi} can be thought of as the deviation from the classical Newtonian regime, quantifying the curvature produced by the acceleration of the expansion of spacetime. A dynamic tracer inside a galaxy cluster will experience two forces:
an inward gravitational force due to the internal mass-energy 
and an additional outward effective force due to the acceleration of the expansion of spacetime.  

Unlike the classical Newtonian regime, in Equation \ref{eq:nabla_phi}, there is a radius relative to a cluster's gravitational center where the acceleration due to gravity balances the acceleration due to the expansion of universe \arnote{\citep{Behroozi+2013}}. This location $r_{\text{eq}}$ is given by $r_{\text{eq}}=(\frac{GM}{-q(z) H^2(z)})^{1/3}$, occurring where $\nabla\Phi=0$. It is a surface that depends on cosmology and the mass internal to it. At \arnote{the redshift of AS1063}, we find $r_{\text{eq}}\sim10\,r_{200}$ for a \arnote{AS1063-size cluster.}


Potentials are only physically meaningful in a relative sense (i.e., escape is to some point as opposed to infinity). 
\arnote{As shown in \citet{miller+2016}, we need to enforce the constraint that a tracer has a relative escape speed of zero at $r_\text{{eq}}$:}
\begin{align}
\label{eq:vesc_final}
    v_{\text{esc}}(r)&=\left[-2(\Psi(r)-\Psi(r_{\text{eq}})) -q(z)H^2(z) (r^2-r^2_{\text{eq}})\right]^{1/2},
\end{align}
\arnote{where the Poisson equation relates the gravitational potential and the density out to $r_\text{{eq}}$}
\begin{equation}
\Psi(r) =-{\rm 4\pi G}\Big{[} \frac{1}{r}\int_{0}^{r}\rho(r')r'^2dr' + \int_{r}^{r_\text{{eq}}}\rho(r')r'dr'\Big{]}.
\label{eq:int_phi_poisson}
\end{equation}

\citet{Behroozi+2013} used $v_{\text{esc}}$ and $r_{\text{eq}}$ to measure the bounded fraction of dark matter particles in simulations. \citet{miller+2016} used $r_{\text{eq}}$ to show that the dark matter and sub-halo escape speed profiles accurately and precisely match expectations from the cluster density profiles. They explored three different analytical density profiles commonly used in the literature: the classic ``NFW'' \citep{navarro+1997}, the ``Einasto'' density profile \citep{einasto+1969}, and the ``Dehnen'' profile \citep{dehen+1993}, for which the two latter profiles both fall off more quickly at large radii compared to the NFW.

\citet{miller+2016} found from simulations that while the Einasto and Dehnen density profiles can accurately predict the escape profile in the phase-space to within $3-5\%$, the NFW profile will tend to overestimate the potential at all radii by $\sim 10\%-15\%$, given the NFW is shallower at larger radii than the Einasto and Dehnen profiles. In this work, we intend to explore these simulation predictions for AS1063 specifically.

\subsection{From theory to observable}
\label{subsec:suppression}
Prior work has focused on using the ``caustic technique'' to infer the mass profile \citep{diaferio+1997}. This technique employs the phase-space data and identifies a number of possible escape edges in the 2D radius/velocity histogram of a cluster phase-space. From these, a velocity ``caustic'' is chosen such that the average of the square of a phase-space isodensity contour is less than or equal to 4 times the square of the velocity dispersion over some radial range. This calibration to the dispersion stems from the virial energy equilibrium of clusters without an accelerating spacetime (see \citet{Gifford2013} for a detailed explanation). In this context, the escape edge is calibrated by the projected velocity dispersion. In order to use it to infer the radial escape edge, \citet{diaferio+1997} and \citet{Diaferio1999}  suggest a geometric model such that:
\begin{equation}
\begin{aligned}
\label{eq:DG1997}
    \langle v_{\text{esc, los}}^2(r_{\perp})\rangle&= 
    \frac{1-\beta_{r_{\perp}}}{3-2\beta_{r_{\perp}}} \langle v_{\text{esc}}^2(r_{\perp})\rangle \\
    & = g^{-1}(\beta_{r_{\perp}}) \langle v_{\text{esc}}^2(r_{\perp})\rangle,
\end{aligned}
\end {equation}
where $\langle \cdot \rangle$ refers to an ensemble average of the galaxies within some radial bin and the measurements are made along lines-of-sight (los) with increasing projected radius $r_{\perp}$. Here, $\beta_{r_{\perp}}$ is the velocity anisotropy measured in projection where
\begin{equation}
\label{eq:beta}
\beta(r) = 1 - \sigma_t^2/\sigma_r^2,
\end{equation}
and $r$ and $t$ refer to radial and tangential components of the velocities.

For the anisotropy-based suppression, typical values of $\beta$ for clusters range from $\beta=0$ to $\beta=0.5$ \citep{Iannuzzi+2012,munari+2013,Stark+2019, Capasso2019}. With this expected level of $\beta$, the observed $v_{\text{esc, los}}$ should be suppressed compared to true $v_{\text{esc}}$ by 50\% - 57\%, regardless of the number of galaxies observed.

Equation \ref{eq:DG1997} has some interesting \arnote{limiting} properties. First, the suppression quickly approaches a constant for increasingly negative values of $\beta$  such that $g^{-1}(\beta \rightarrow -\infty) \sim 0.7$. In other words, all projected phase-space escape edges must be suppressed by at least 30\% compared to the 3D edge. Second, as the anisotropy becomes radial ($\beta \rightarrow 1$), the suppression grows to $g^{-1}(\beta) \rightarrow 0$, such that no edge would be detectable in projected data when the tracers are dominated by motions corresponding to radial infall. Third, for isotropic systems ($\beta =0$), the suppression is $g^{-1}(\beta) = 1/\sqrt{3}$ \citep{halenka+2020}.


\arnote{The number of tracers does not appear in equation \ref{eq:DG1997}. Yet \citet{Serra2011} noticed a dependence on the number of tracers in simulation phase-spaces. Motivated by this issue and prior work by \citet{Gifford_2017}, } \citet{halenka+2020} took another approach. They recognized that there is always the possibility that a tracer is observed escaping along the line-of-sight. As one fills the phase-space with additional tracers, more such galaxies should exist. They proposed that the suppression in observed escape edges should simply be due to low statistical sampling of the phase-space. \citet{halenka+2020} utilized a purely analytical approach to sample cluster phase-spaces via action-based modeling \citep{Vasiliev2019}. They show that in an idealistic scenario where hundreds of thousands of tracers are projected in velocity and radius, the true escape profile can be accurately observed in projected data. They then show quantitatively that the suppression of the escape edge is dominated by statistical sampling. Anisotropy, cosmology, and cluster mass play very minor roles (e.g., a maximum of a few percent).

\citet{halenka+2020} provide a model of the suppression of the projected escape profile :
\begin{align}
\langle v_{\text{esc, los}}\rangle(r_{\perp})&= \frac{\langle v_{\text{esc}}\rangle(r_{\perp})}{Z_v(N)}
\\
\label{eq:Zv}
    Z_v(N)&=1+\left(\frac{N_0}{N}\right)^{\lambda},
\end{align}
where $N$ is the number of tracers in the projected phase-space between $0.3\,r_{200} \le r \le r_{200}$. \arnote{$N_0$ and $\lambda$ are different depending on the data used to calibrate the suppression. For instance, the Action-based Galaxy Modeling Architecture (AGAMA) \citep{Vasiliev2019} can be used to generate analytic mock phase spaces. Similarly, semi-analytic galaxies from massive halos in the Millennium N-body can also be employed. Both datasets produce phase-spaces which include interlopers. Galaxies with large 3D distances can sometimes be projected onto smaller radii on the sky plane. The Millennium galaxies also include realistic sub-structure as well as a cosmological background of galaxies. While the suppression values inferred from the two datasets are statistically the same, there are differences in their means we might be able to constrain observationally. }


\arnote{At the same time, two mechanisms have been proposed to explain the suppression of escape edge profiles: Equation \ref{eq:DG1997} and Equation \ref{eq:Zv}. These have fundamentally different physical explanations and we aim to provide the first observed value for the suppression and address the validity of the models which can explain it.}


\arnote{
\section{AS1063 Phase-Space Edge Analysis}
\label{sec:Edge measurement}}

\arnote{
\subsection{Inference of the Phase-Space Edge}
\label{sec:membership}}

\arnote{
\subsubsection{Assignment of Projected Radii and Velocities}
\label{subsubsec:Assignment of Projected Radii and Velocities}}

\begin{figure}
\includegraphics[width=.5\textwidth]{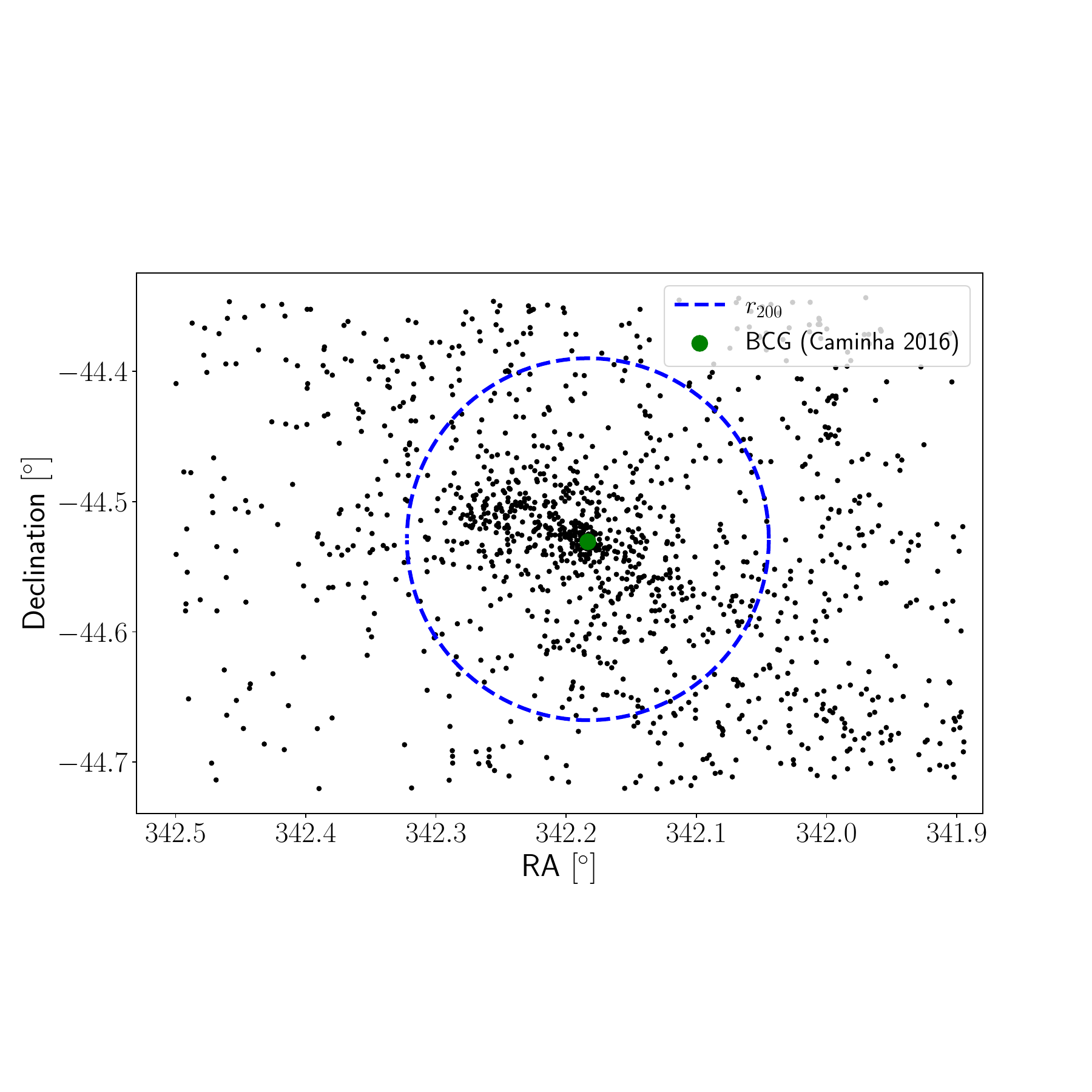}
\caption{\label{fig:phase_space_interlopers} Distribution of the 1305 redshifts for AS1063 on the sky (black points).   The green point indicates the position of the BCG ([R.A.=$22^{\text{h}}48^{\text{m}} 44.0^{\text{s}}$, Dec.$=-44^{\text{d}}31^{\text{m}} 51.0^{\text{s}}$]) \citep{caminha+2016}. These are shown relative to $r_{200}=2.63\,\text{Mpc}$ (the blue, dashed, line), obtained from \citet{sartoris+2020}.
}
\end{figure}

We use data for AS1063 from CLASH-VLT observations, as part of the Cluster Lensing And Supernova survey with Hubble program \citep{Postman+2012,lotz+2017}. The galaxy spectra we use are from \citet{karman+2015,mercurio+2021}, and the full dataset contains 3850 redshifts. Following \citet{mercurio+2021}, we apply a redshift cut between $0.30905 <z< 0.3716$ as a starting point for the phase-space analysis. We do not apply any additional radial cuts, although all galaxies after this initial redshift cut are within $r_{\perp}<5\,\text{Mpc}$. The above cuts yield 1305 galaxies remaining in the sample.

\arnote{To obtain the AS1063 phase-space}, we start by assigning line-of-sight peculiar velocities to each galaxy, via
\begin{align}
\label{eq:los}
    v_{\text{los}}&=c\frac{z_g-z_c}{1+z_c},
\end{align}
where $z_g$ is the galaxy redshift, $z_c$ is the mean cluster redshift, and $c$ is the speed of light. The projected radius for each galaxy is calculated for our chosen cosmology and the galaxy redshifts:
\begin{align}
\label{eq:r_perp}
    r_{\perp}&=r_\theta\left(\frac{1}{1+z_c}\frac{c}{H_0}\int_{0}^{z_g}\frac{dz'}{E(z')}\right),
\end{align}
where $r_\theta$ and $r_\perp$ are the angular and projected physical separation between the galaxy and the center of the cluster, and $E(z)=\left[\Omega_\Lambda +\Omega_M (1+z)^3\right]^{1/2}$ for a flat $\Lambda \text{CDM}$ universe. \arnote{We adopt redshift and spatial centers of $z_c=0.3452$ and [R.A.=$22^{\text{h}}48^{\text{m}} 44^{\text{s}}$ (about $13"$ offset from the BCG), Dec.$=-44^{\text{d}}31^{\text{m}} 44^{\text{s}}$] respectively, measured from galaxies solely within the virial radius of $r_{200} = 2.4$ Mpc, given the \citet{sartoris+2020} lensing estimate.}

\arnote{With these projected radii and velocities}, we cull all galaxies with velocities $>|4500|$ km/s as these are readily identified as non-cluster members. \arnote{Additionally, we restrict projected radii to only be only within $0.2 < r_{\perp}/R_{200} < 1$, where the spectroscopic completeness of the sample is approximately uniform \citep{mercurio+2021}, yielding 645 galaxies.}
\vspace{0.5cm}
\arnote{
\subsubsection{Interloper Analysis}
\label{subsubsec:Interloper Analysis}}
\citet{fadda+1996} showed that in phase-space data there can be ''interlopers'' that exist outside the main body of the cluster. These are identifiable based on velocity gaps in the data. Like many others, we use a shifting-gapper technique \citep{fadda+1996,girardi+1996,adami+1997, Wing2013} to identify these interlopers in order to infer the phase-space edge profile. We then choose a velocity gap size and a radial bin size to conduct the shifting--gapper technique on the remaining galaxies and identify interlopers to remove. 

\arnote{In deciding the value for the gap and the binning scheme, an examination of the literature shows conflicting choices in the gap size, with 500km/s or 1000km/s being the most common choices \citep{Fadda1996,Wing2013,Sifon2013, Gifford2013, Crawford2014,Barsanti2016,Sifon2016}. In terms of binning, the typical choices is between 10 and 30 galaxies per bin. We note that \citet{sartoris+2020} used an 800km/s gap and a minimum of 15 galaxies per bin to define interlopers in the first step of their membership measurement for AS1063. }

\begin{figure}
\includegraphics[width=.5\textwidth]{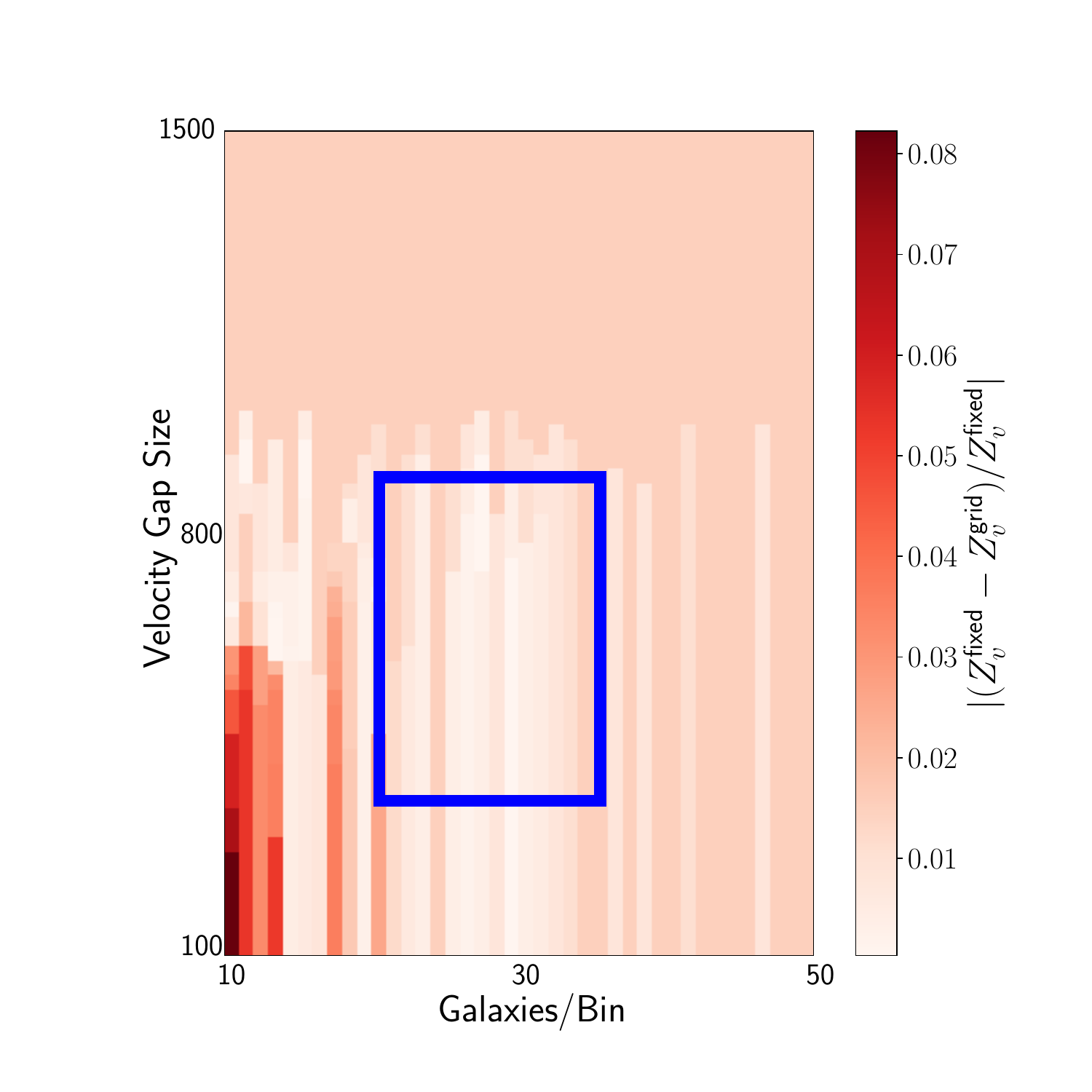}
\caption{\label{fig:Shifting_gapper_theory_test} 
\arnote{Fractional differences between edges of Millennium halos with variable bin/GAP parameters, and fixed shifting-gapper parameters (the edge itself is divided into 5 total bins, while the shifting-gapper parameters are 15 galaxies per bin and a velocity gap of 800 km/s, identical to \citet{halenka+2020}). Halos are enforced to have at least 400 galaxies within $0.3 < r_{\perp}/R_{200} < 1$, and the fractional differences above are averaged over the 5/100 halos in the sample which satisfy this criterion. We measure a statistically negligible difference ($\sim 0.5\%$) between the variable and fixed bin/GAP edge definitions, indicating that the broadening of the interloper range shown as opposed to fixed parameters does not induce a source of bias on the edge relative to \citet{halenka+2020}}}
\end{figure}

\begin{figure*}
\includegraphics[width=.5\textwidth]{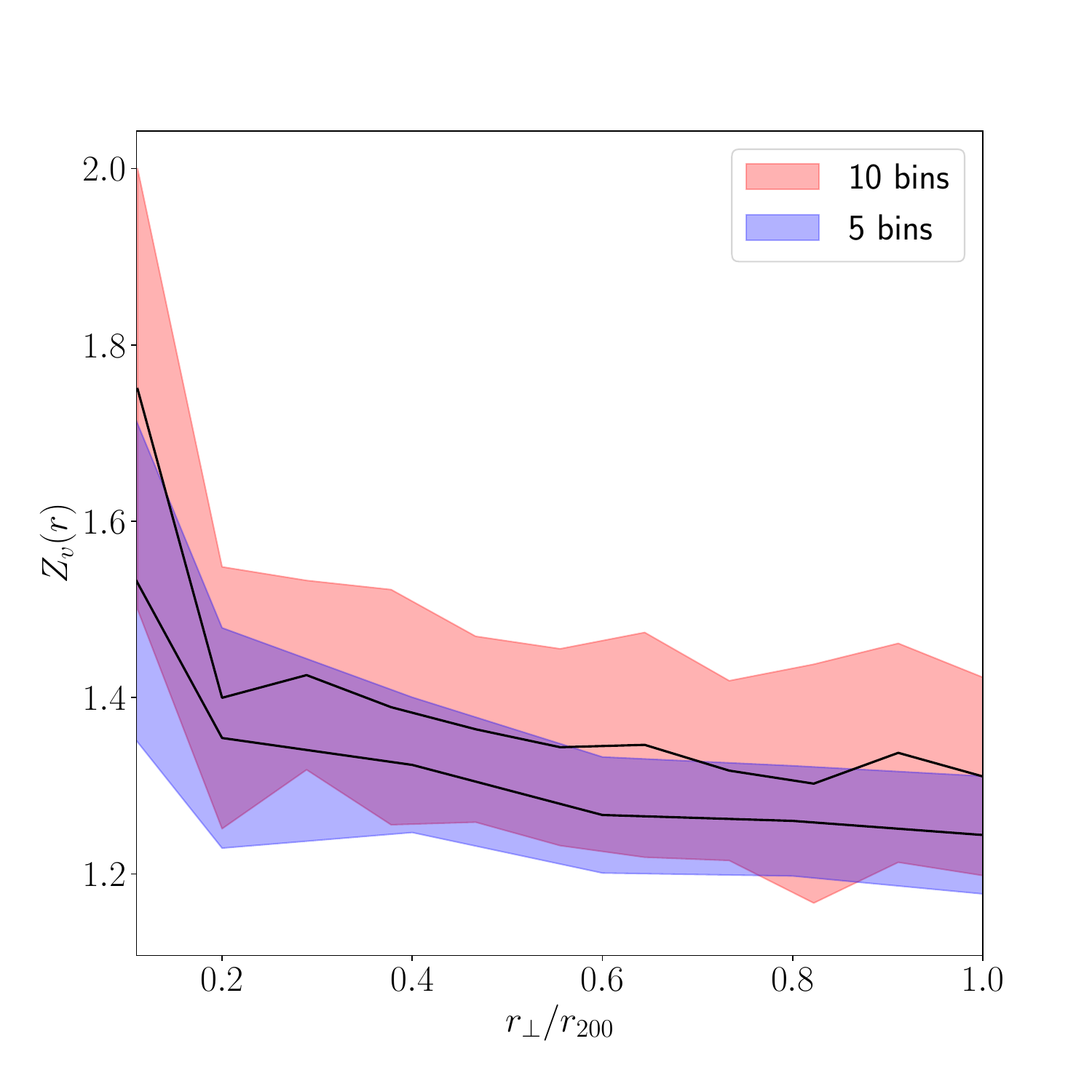}
\includegraphics[width=.5\textwidth]{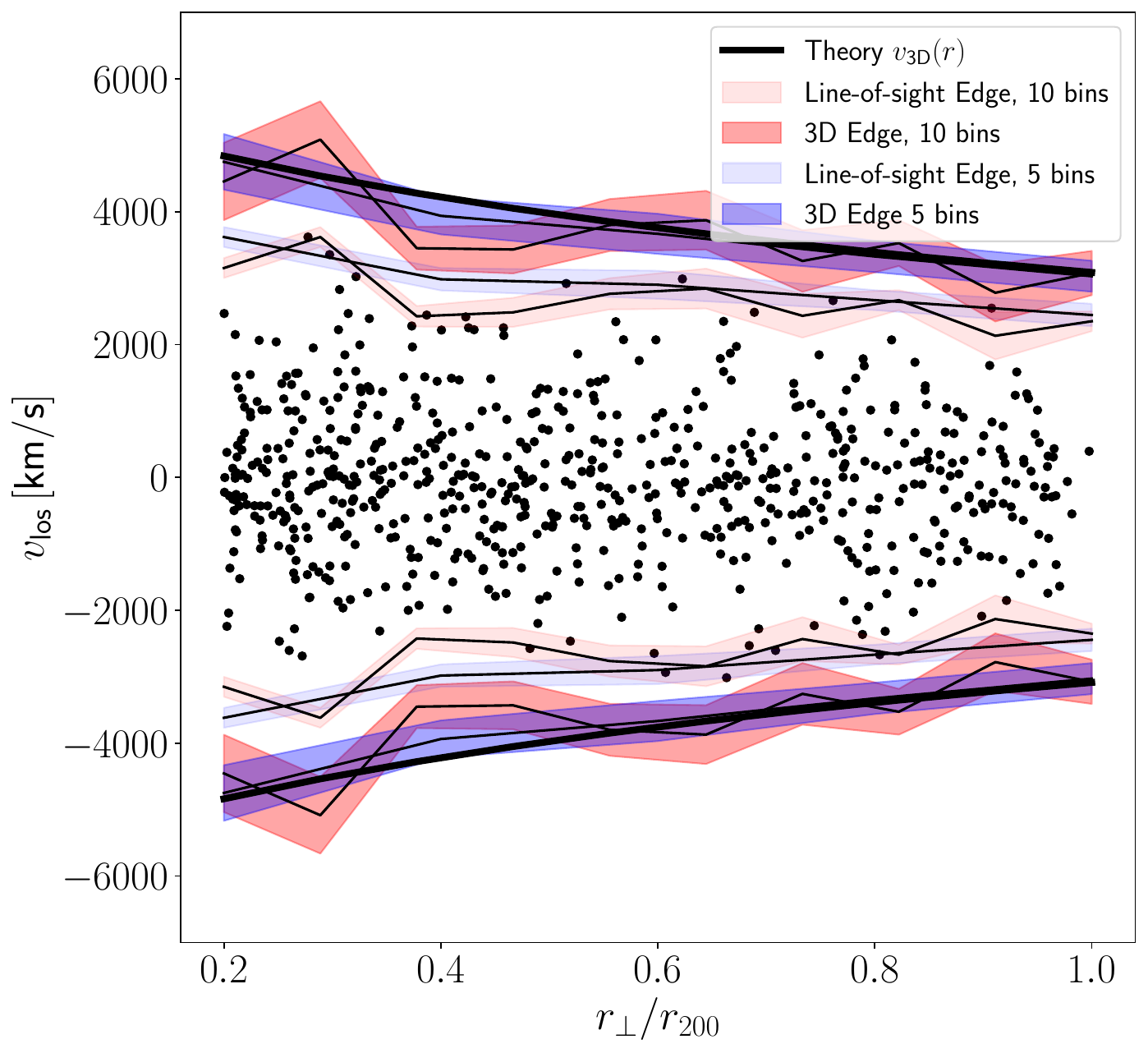}
\caption{\label{fig:Zv_use} 
\arnote{$\textit{Left panel}$: Predicted radial suppression of a AS1063-type cluster using the AGAMA analytic model.  A smaller bin-width corresponds to lower sampling per bin and hence more suppression, as shown by the fact that $Z_v$ is higher for 10 bins than 5 bins.  $\textit{Right panel}$: Example analytic phase-space of a AS1063-type cluster using AGAMA, where the edge is again measured  using 5 (red regions) and 10 bins (blue regions). It is clear that a lower bin width yields more suppression, 
however the same 3D escape profile is still recovered (to within uncertainty) so long as the correct bin calibration is used.}}
\end{figure*}

\begin{figure}
\includegraphics[width=.5\textwidth]{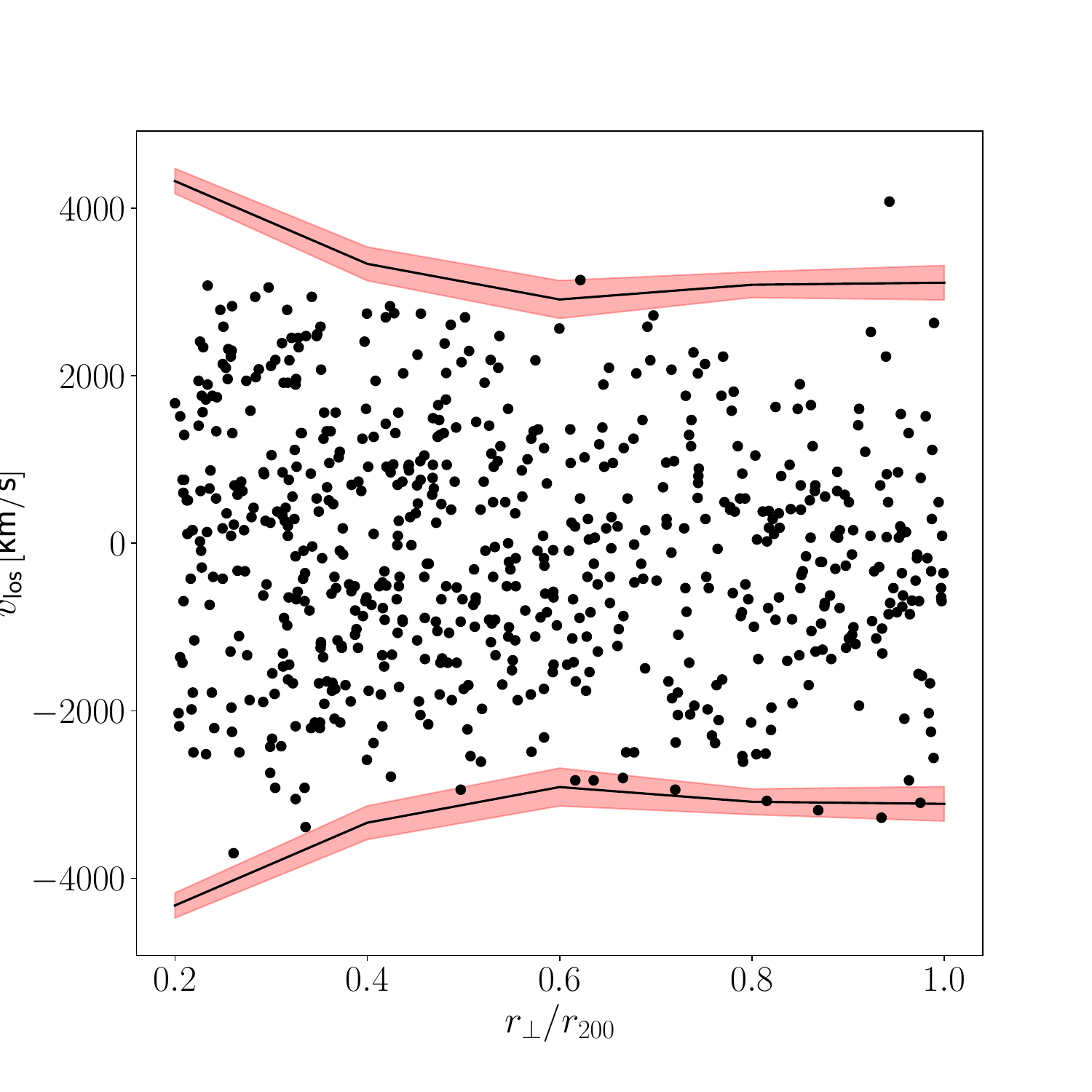}
\caption{\label{fig:Edge_updated} 
\arnote{The phase-space edge for Abell S1063 inferred from the galaxy data after radial and velocity cuts from $0.2\,r_{200}<r_{\perp}<r_{200}$ (the range where the suppression function is used, and also spectroscopic completeness is approximately uniform). Here, the edge incorporates the scatter (red bands) from interloper variation}.}
\end{figure}

Different bin sizes and velocity gaps will often identify different interlopers, and there is no obvious ``best'' {\it{a priori}} parameter choice. \arnote{\citet{Old2014} compared two values for the gap size in clusters from an N-body simulation, and found that the mass inferred using the velocity dispersion using the smaller gap size had a smaller mass than the cluster with the larger gap. The clusters with the smaller gap size were also more discrepant from the true $M_{200}$ compared to the large gap size. \citet{Crawford2014} performed a systematic study by varying the number of galaxies per bin as well as the gap size. They found a weak trend of an increasing mass with higher galaxies per bin, and a stronger trend of increasing mass with higher velocity gap size.} 

\arnote{Similar to \citet{Crawford2014}, we vary the bin size and velocity gap to identify interlopers and compare the measurement to a fiducial one. While they studied the dispersion, we study the edge suppression ($Z_v$). For a baseline constraint on accuracy and precision, we use clusters from N-body simulations where the true edge and suppression is known.  We use the halo sample from \citet{halenka+2020} which was based on semi-analytic galaxies in the Millennium simulation. We use their galaxy data having the dimmer apparent magnitude limit for semi-analytic galaxies to maximize the phase-space sampling.  We then choose the five halos with the highest sampling and similar to AS1063 with $N_\text{{gal}} >500$ within $0.2 < r_{\perp}/R_{200} < 1$. This exercise could be repeated for halos with lower sampling, but that is beyond our scope.}

\arnote{Using projected phase-spaces, we then measure the escape edges over a multiple lines-of-sight for a fixed set of shifting-gapper parameters. We use the same parameters of 800 km/s and 15 galaxies/bin from \citet{halenka+2020}. Using the underlying 3D edges, we recover the radially averaged suppression from \citet{halenka+2020} which we label as $Z_v^{\rm fixed}$. We then measure the edge over a wide range of shifting-gapper parameters and calculate new suppression values, one for each of the $(i,j)$ pairs of bin and gap sizes ranging from 10-50 galaxies/bin and a velocity gap ranging from 100-1500 km/s. The accuracy of these edges can then be assessed by calculating the mean and standard deviation of the absolute value of the fractional difference, $| f (\Delta Z_v)|$, between $Z_v^{\rm{fid}}$ and $Z_v^{i,j}$}.

\arnote{These fractional differences are shown in Figure \ref{fig:Shifting_gapper_theory_test}. There is a clear high variance region below 20 galaxies/bin. Above a velocity gap of $\sim 1000$ km/s and $\sim 35$ galaxies/bin, we see convergence to $ |f (\Delta Z_v)| \sim 0.02$, i.e. that the edges become biased relative to the fixed case. Given a redshift uncertainty of 150 km/s \citep{mercurio+2021}, we identify a velocity gap floor of twice this uncertainty or 300 km/s. The upper bound of the velocity gap is chosen to keep $| f (\Delta Z_v)|$ low, where 800 km/s is the maximum gap where $| f (\Delta Z_v)|<1\%$. This corresponding parameter space where the edge measurement is robust to velocity gap and bin variation is shown by the blue box in Figure \ref{fig:Shifting_gapper_theory_test}, where we measure $ |f (\Delta Z_v)| = 0.005 \pm{0.003}$. }

\arnote{For a system with observed data like AS1063, we have used simulation halos to show that, for a wide range of gap-size/bin choices, the systematics induced into the inferred 3D edge from the projected data are at the sub-percent level. We also use the simulations to constrain the measured edge scatter, which we carry through the rest of the analyses.}


\arnote{\subsection{Binning and Smoothing the Edge Profile}}
\label{subsec:binning}

\arnote{In order to calculate the phase-space maxima, we need to choose a bin size. Typically, a nominal physical bin size is chosen, e.g., 0.2 Mpc \citep{Stark+2019}. However, prior characterizations of cluster escape profiles have used data with significantly lower sampling. In this section, we explore the effect of bin size to make an informed as opposed to arbitrary choice for the bin size.}

\arnote{We explore the bin size using the Action-based Galaxy Modeling Architecture (AGAMA) framework \citep{Vasiliev2019}. Distinct from N-body simulations such as Millennium, This technique relies on forward modeling a cluster using Action/Angle variables \citep{binney+2008}.  Unlike a Jean's inversion scheme (e.g. \citet{sartoris+2020}), AGAMA uses a set of predefined characteristics of a cluster (e.g. the mass, density profile, velocity anisotropy, cosmology, etc.) in order to see how these affect the phase-space characteristics. In our analysis of the analytic halos, we use the same mass (assuming a Dehnen density profile) and sampling ($\sim 800$ tracers within $r_{200}$) as AS1063 (although note that \citet{halenka+2020} found the suppression is not affected by the mass of the system). AGAMA is preferable to Millennium in this context given the systematic control of parameters it allows us.} 

\arnote{To incorporate the effects of cosmology, we cull all tracers outside the cosmological escape profile (Equation \ref{eq:vesc_final}) for the pre-specified cosmology in~\S\ref{sec:Introduction}. As mentioned in~\S\ref{sec:Introduction}, AGAMA however lacks a cosmological background as well as realistic substructure. We trade these effects however for systematic control of phase-space characteristics. In the projected analytic phase-spaces, we follow the same projection procedure from \citet{Gifford2013,halenka+2020} and treat the viewing angle as a random variable along the z-axis. We simulate 100 different realizations of phase-spaces for each mock AGAMA cluster. The clusters are placed at the same redshift as AS1063, for which we sample 100 different viewing angles.}

\arnote{In the left panel of Figure \ref{fig:Zv_use},
we show how the phase-space edge profile differs for different bin choices, corresponding to 5 bins (blue shaded region) and 10 bins (red shaded region). Here, the suppression is measured for the phase-space edge between $0.2\,r_{200}<r_{\perp}<r_{200}$ where 100 different line-of-sight draws are performed. It is apparent that the higher bin count yields more suppression at all radii, which can be attributed to the fact that the average sampling per bin is reduced for a smaller bin width. }

\arnote{In the right panel of Figure \ref{fig:Zv_use}, we show the effect on the edges from the number of bins. This is for a single line-of-sight draw on a single projected AGAMA phase-space. The edge from 10 bins is more suppressed than the edge from 5 bins. However, so long as this increased suppression is accounted for using the analogous $Z_v$ (e.g., from the left panel of Figure \ref{fig:Zv_use}), the 3D edge profile will still be inferred. The final accuracy and precision on the edge from binning is contained within the $Z_v$ function. In addition to the scatter from gapsize/bin-size in Section \ref{subsubsec:Interloper Analysis}, we incorporate the suppression uncertainty in the mass inference.}

\arnote{For AS1063, we will use 5 bins for the rest of this paper and the appropriate $Z_v$ from Figure \ref{fig:Zv_use}. Furthermore, we include the radial shape dependence of the suppression function, which is higher in the core compared to $r_{200}$. Specifically, we have used the AGAMA framework to calibrate $Z_v(r)$ for 5 bins and for a sample of galaxies selected within the range $0.2\,r_{200}<r_{\perp}<r_{200}$  where the spectroscopic completeness function is approximately uniform from \citep{mercurio+2021}.}



While potentials are smooth on the scales probed here, the measured edges can be noisy. There are various techniques used to control this effect \citep{Serra2011, Gifford_2013}. In our work, we employ smoothing functions, which of course relates closely to binning. Our smoothing analysis requires us to choose the number of bins and also a form for the parameterized function to smooth the raw data. We then measure the statistical bias and variance. In this context, the square of the bias of the model quantifies the degree to which the model is overfit, while the variance of the model quantifies the degree to which the model is underfit. If these quantities are equal, or their sum is minimized, the fit will be optimized, and can be quantified according to the statistical risk, or mean square error (MSE), of the model, given by \citep{keener2010}
\begin{align}
\label{eq:risk}
    \text{risk}&=(\theta-\mathbb{E}[\delta])^2+\mathbb{E}[(\delta-\mathbb{E}[\delta])^2],
\end{align}
Here, $\mathbb{E}[x]$ is the expected value of the variable $x$, $\theta$ is the escape edge, and $\delta$ is our estimator we compare the model to, or an NFW fit of the escape profile, obtained through MCMC (see~\S\ref{subsec:mass_estimation}). 

\arnote{To measure the MSE, we need to choose a function to make the estimator $\delta$. We explore a range of different smoothing functions, with polynomial degrees from one through five (linear, quadratic, etc.). We find quadratic and cubic smoothing functions minimize the MSE by $\sim70\%$ relative to the unsmoothed edge and higher order polynomials. Hence, we choose to use these fits to avoid over and under-fitting of the edge profile\footnote{\arnote{It is worth noting that the virial mass is statistically unaffected by the choice of smoothing. Rather, smoothing has the strongest  effect on the edge profile near the core, which primarily controls the measured concentration.}}.}

\vspace{0.5cm}

\section{Mass Estimation}
\label{subsec:mass_estimation}
\arnote{
\subsection{From the Phase-Space Edge to $M_{200}$}}

\arnote{The phase-space edge for AS1063 is shown in Figure \ref{fig:Edge_updated}, where the errorbars combine interloper and redshift error. To model the effective potential profile from this edge, we use the NFW potential \citep{miller+2016}
\begin{align}
\label{eq:nfw_psi}
    \Psi(r)&=-\frac{4\pi G\rho_s r_s^2 \log(1+r/r_s)}{r/r_s},
\end{align}
where $r_s$ is the scale radius and $\rho_s$ is the normalization. However in modeling the effective potential profile, we need to account for the phase-space edge suppression due to sampling, and also over and under-fitting. We use the $Z_v(r)$ in Figure \ref{fig:Zv_use} to account for suppression, and a quadratic smoothing function to account for over and under-fitting}.

\arnote{Given the 3D phase-space edge (including its error from suppression), we use equations \ref{eq:vesc_final} and \ref{eq:Zv} to infer the mass profile using the Poisson equation. We use Bayes' theorem and Goodman \& Weare’s affine invariant Markov chain Monte Carlo (MCMC) Ensemble sampler to model the posteriors of the NFW density profile \citep{emcee}, corresponding. In all cases we use a Gaussian likelihood, where the error comes from the phase-space edge uncertainty. To find $M_{200}$, we interpolate the cumulative mass density to identify the radius which is 200$\times$ the critical (and also mean) density and report the corresponding mass within that radius as $M_{200}$.}

\arnote{We consider two fitting options to the NFW: a single parameter model using two different mass-concentration relations and a two parameter model fitting to the central density and scale radius. The priors are summarized in Table \ref{table:Summaray_priors}. Here, we present only the NFW profile, while the other profiles are presented in the Appendix. }

\arnote{In the top left panel of Figure \ref{fig:phase_space}, we show the 2D posterior  for the NFW profile's parameters.  We use the GTC package \citep{Bocquet+2016} to generate the histograms. The covariance of $r_s$/$\rho_s$ suggests that only a slim range of these parameters yield physical mass constraints, both in cases when the edge profile is smoothed and unsmoothed. The top right panel converts our  $r_s$/$\rho_s$ posterior to $M_{200}/c_{200}$, measured with respect to the mean density. We find that when smoothing the edge profile, our $M_{200}/c_{200}$ is consistent with \citet{gruen2013}, shown by the blue contours. If no smoothing functions are used, the mass profile is relatively unchanged, but the concentration increases by $\sim0.5\sigma$. Further, the covariance in $M_{200}/c_{200}$ yields a different shape when the escape velocity is used as opposed to lensing, where we find the mass varies hardly at all with the concentration. This is expected from the fact that the potential profile is a second derivative of the density profile and therefore much flatter so that it is less well constrained by the data. This analysis suggests that future joint analyses using lensing and the escape profiles could place tight constraints on the mass-concentration relation.} 

\arnote{In the bottom panel of Figure \ref{fig:phase_space}, we show the escape profile inferred from the smoothed two-parameter (M$_{200}$, c$_{200}$) model, with respect to the critical density. The agreement with \citet{gruen2013} is highlighted by the agreement in our escape profile and their inferred escape profile, given their $M_{200}/c_{200}$ posterior.
While more recent lensing estimates are available (see Table \ref{table:Summaray_mass_estimates}), we use \citet{gruen2013} as a baseline given their published mass/concentration posterior contours. }

\arnote{The results (16.5,50,83.5 percentiles of the 1D marginalized posteriors) for all of our the NFW fits (in addition to the Einasto and Dehnen fits from the Appendix) are provided in Table \ref{table:mass_estimates_this_work}.}

\begin{figure*}[ht]
\includegraphics[width=.31\textwidth]{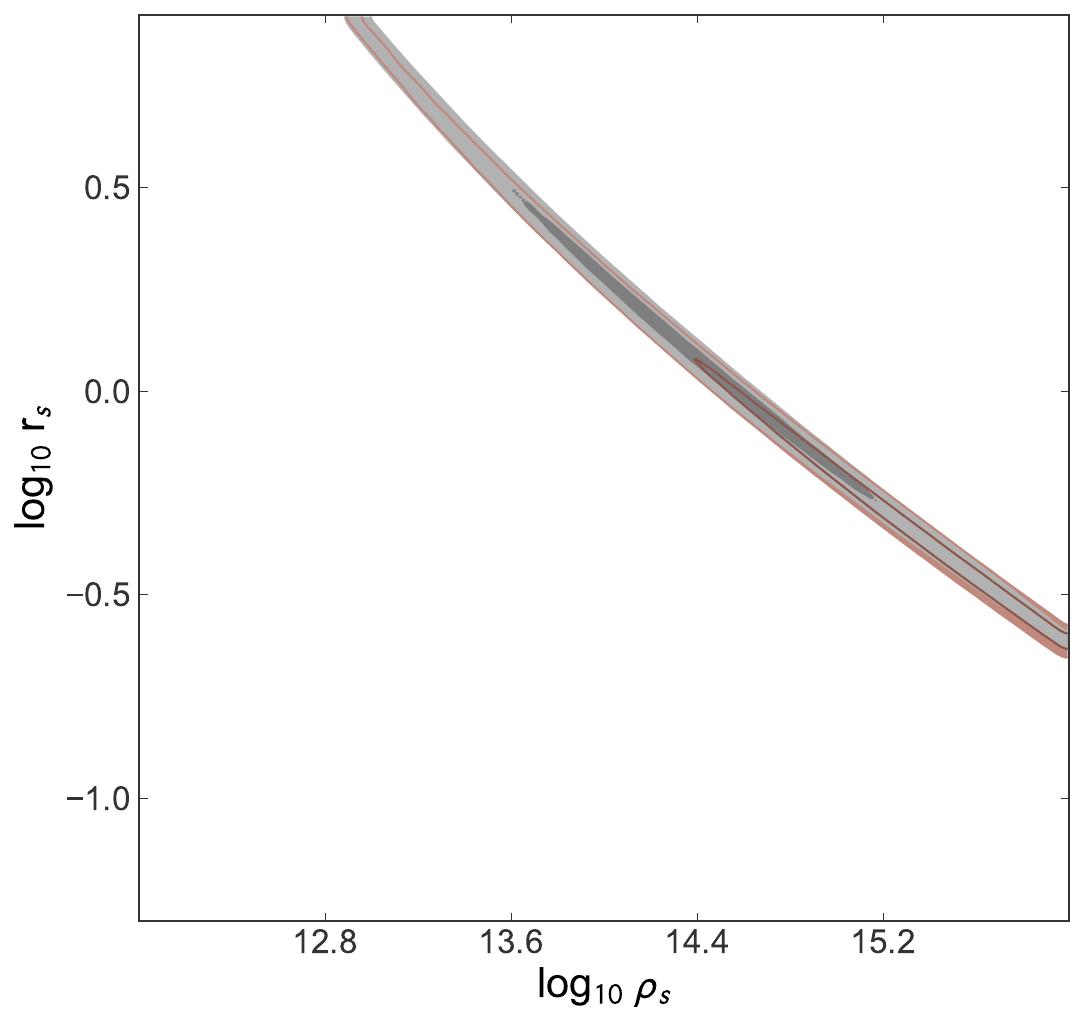}
\includegraphics[width=.35\textwidth]{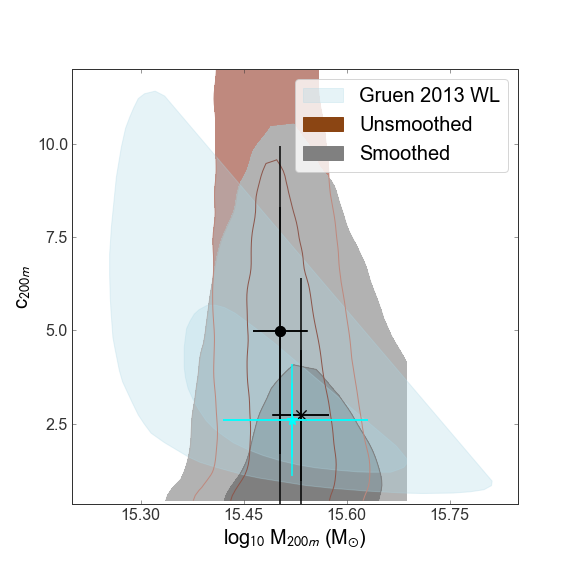}
\includegraphics[width=.32\textwidth]{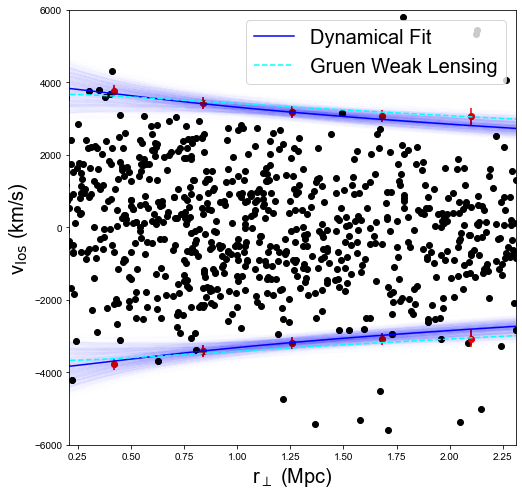}
\caption{\label{fig:phase_space} \arnote{$\textit{Left panel}$: 2D posterior estimates  (68\% and 95\%) of the NFW profile parameters. The covariance of the distribution suggests that only a narrow range of $r_s$ and $\rho_s$ values yield physical mass constraints. $\textit{Right panel}$: Conversion of the $r_s$ and $\rho_s$ posterior to the concentration and mass, measured with respect to the mean density of the universe. We show the posteriors for both the smoothed and unsmoothed edges, where it is apparent smoothing primarily controls the concentration. The blue contours are taken from the weak-lensing analysis of \citet{gruen2013}. The covariance in mass/concentration has a different shape than that from weak lensing, where our mass posterior is nearly independent of concentration. $\textit{Right panel}$: The best-fit two parameter (M$_{200}$, c$_{200}$) NFW escape profile relative to the \citet{gruen2013} escape profile, inferred from their mass. These profiles are shown relative to all the galaxy data (including interlopers).}}

\end{figure*}

\begin{table*}
\caption{AS1063 Potential Profile and Mass Constraints. Central values are medians, while errors are 16.5\% and 83.5\% percentiles of the marginalized 1D posteriors. }
\label{table:mass_estimates_this_work}
\hspace*{-0.5in}
\begin{tabular}{cccccccl}
    \multicolumn{6}{c}{NFW 1 parameter} \\ \cline{1-7}
Edge & log$_{\rm 10}$ M$_{200}$  & & & & & & Notes \\ \hline

unsmoothed &   \arnote{15.39$^{+0.03}_{-0.04}$} &  & & & & & \citet{duffy+2008}\\ 

smoothed   &   \arnote{15.40$^{+0.03}_{-0.03}$ }&  & & & & & - \\ 
 
unsmoothed   &   \arnote{15.59$^{+0.04}_{-0.04}$}        & & & & & & $200\times \rho_{{\rm mean}}$) \\ 

smoothed   & \arnote{15.60$^{+0.04}_{-0.04}$} & & & & & & -\\   \\
   
    \multicolumn{6}{c}{NFW 2 parameter} \\ \cline{1-7}
Edge & log$_{\rm 10}$ M$_{200}$& c$_{200}$            & r$_{200}$            & log$_{\rm 10}$ $\rho_s$ & log$_{\rm 10}r_s$ & & Notes \\ \hline

 unsmoothed & 15.42$^{+0.06}_{\arnote{-0.09}}$ & \arnote{3.53$^{+3.81}_{-2.43}$} & \arnote{2.50$^{+0.10}_{-0.16}$}  &  \arnote{14.89$^{+0.73}_{-0.99}$} &  \arnote{-0.14$^{+0.48}_{-0.31}$} &  &
 fit to $\rho_s$, r$_s$  \\

  smoothed & 
  \arnote{15.40$^{+0.06}_{-0.12}$} & \arnote{1.91$^{+2.8}_{-1.31}$} & \arnote{2.46$^{+0.12}_{-0.2}$}  & 14.34$^{+0.86}_{-0.88}$ &  -0.12$^{+0.46}_{-0.39}$  & -\\

unsmoothed & 15.54$^{\arnote{+0.04}}_{-0.04}$ & \arnote{4.94$^{+5.07}_{-3.28}$}  & \arnote{3.52$^{+0.10}_{-0.10}$}   &  
&  & &  $200\times \rho_{{\rm mean}}$\\

  smoothed & 
15.55$^{\arnote{+0.04}}_{\arnote{-0.04}}$ & 
\arnote{2.77$^{+3.75}_{-1.81}$} & \arnote{3.57$^{+0.10}_{-0.10}$} & 
&  & &  -\\  
  
  \\

    \multicolumn{8}{c}{Einasto 3 parameter} \\ \cline{1-7}
Edge &    log$_{\rm 10}$ M$_{200}$& c$_{200}(r_{-2})$            & r$_{200}$            & log$_{\rm 10}$ $\rho_0$ &   log$_{\rm 10}$h                  & n & Notes \\ \hline
unsmoothed & 
\arnote{15.40$^{+0.08}_{-0.16}$} & \arnote{3.79$^{+5.15}_{-2.67}$}  & \arnote{2.46$^{+0.15}_{-0.29}$}   &   \arnote{19.25$^{+1.62}_{-1.85}$}  & 
\arnote{-6.42 $^{+2.85}_{-2.23}$} & \arnote{5.82$^{+1.55}_{-1.99}$} & 
fit to $\rho_0$, h, n \\ 

smoothed & 
\arnote{15.34$^{+0.11}_{-0.23}$} & \arnote{1.95$^{+2.86}_{-1.55}$} & \arnote{2.35$^{+0.21}_{-0.38}$} &   \arnote{18.81$^{+1.58}_{-1.72}$}  & 
\arnote{-6.50$^{+2.82}_{-2.37}$} & \arnote{6.11}$^{+1.55}_{\arnote{-2.03}}$ & - \\

unsmoothed &   
\arnote{$15.54^{+0.08}_{-0.16}$} & \arnote{$3.79^{+5.15}_{-2.67}$}  & \arnote{$2.46^{+0.15}_{-0.29}$} & 
&  &  & $200\times \rho_{{\rm mean}}$\\

smoothed &   
\arnote{15.49}$^{+0.07}_{\arnote{-0.12}}$ &
\arnote{$2.81^{+3.83}_{-2.17}$}  & \arnote{$3.38^{+0.17}_{-0.30}$} & 
&  &  & - \\
    \\

    \multicolumn{8}{c}{Dehnen 3 parameter} \\ \cline{1-7}
 Edge &   log$_{\rm 10}$ M$_{200}$&          & log$_{\rm 10}$ r$_{200}$            & log$_{\rm 10}$ $M_{{\rm tot}}$ &   r$_s$                  & $\gamma$ & Notes \\ \hline

unsmoothed &   
\arnote{15.43}$^{+0.05}_{\arnote{-0.10}}$ & 
 &
 \arnote{2.52}$^{\arnote{+0.10}}_{-0.10}$  &   \arnote{$15.88^{+0.25}_{-0.18}$}  & 17.27 $^{+8.63}_{-9.68}$ & 2.09$^{+0.12}_{-0.19}$ & fit to M$_{tot}$, r$_s$, $\gamma$ \\ 

smoothed & 
\arnote{15.39}$^{+0.08}_{\arnote{-0.17}}$ &  & \arnote{2.45}$^{\arnote{+0.08}}_{-0.17}$ &   \arnote{$16.03^{+0.28}_{-0.20}$}  & 14.76 $^{+9.99}_{-9.22}$ & 1.74$^{+0.19}_{-0.36}$ &  -\\

unsmoothed &   
\arnote{15.51}$^{\arnote{+0.04}}_{-0.05}$ &  & 
3.44$^{\arnote{+0.11}}_{\arnote{-0.13}}$ &   &  &  & $200\times \rho_{{\rm mean}}$\\

smoothed &  
\arnote{$15.51^{+0.04}_{-0.08}$} &   & \arnote{$3.54^{+0.13}_{-0.18}$} &  &  &  & - \\
    \\

    \end{tabular}
 \end{table*}

\subsection{Dependence on the Density Profile}
\label{subsec:Dependence on the Density Profile}

\citet{miller+2016} noticed that when constraining against the NFW density profile of dark matter halos, the inferred escape velocity was higher than what was actually observed in simulations. They attribute this to the fact that the NFW density profile is too shallow in the outskirts for realistic DM halos (see also \citet{Diemand+2004}). In our results, we are not constraining to the density, but instead to the escape edge. By definition, the NFW, Einasto, and Dehnen potential profiles will be nearly identical to each other within the fitting errors, due to the fact that they are being fit to the same edge. 

\begin{figure}
\includegraphics[width=.5\textwidth]{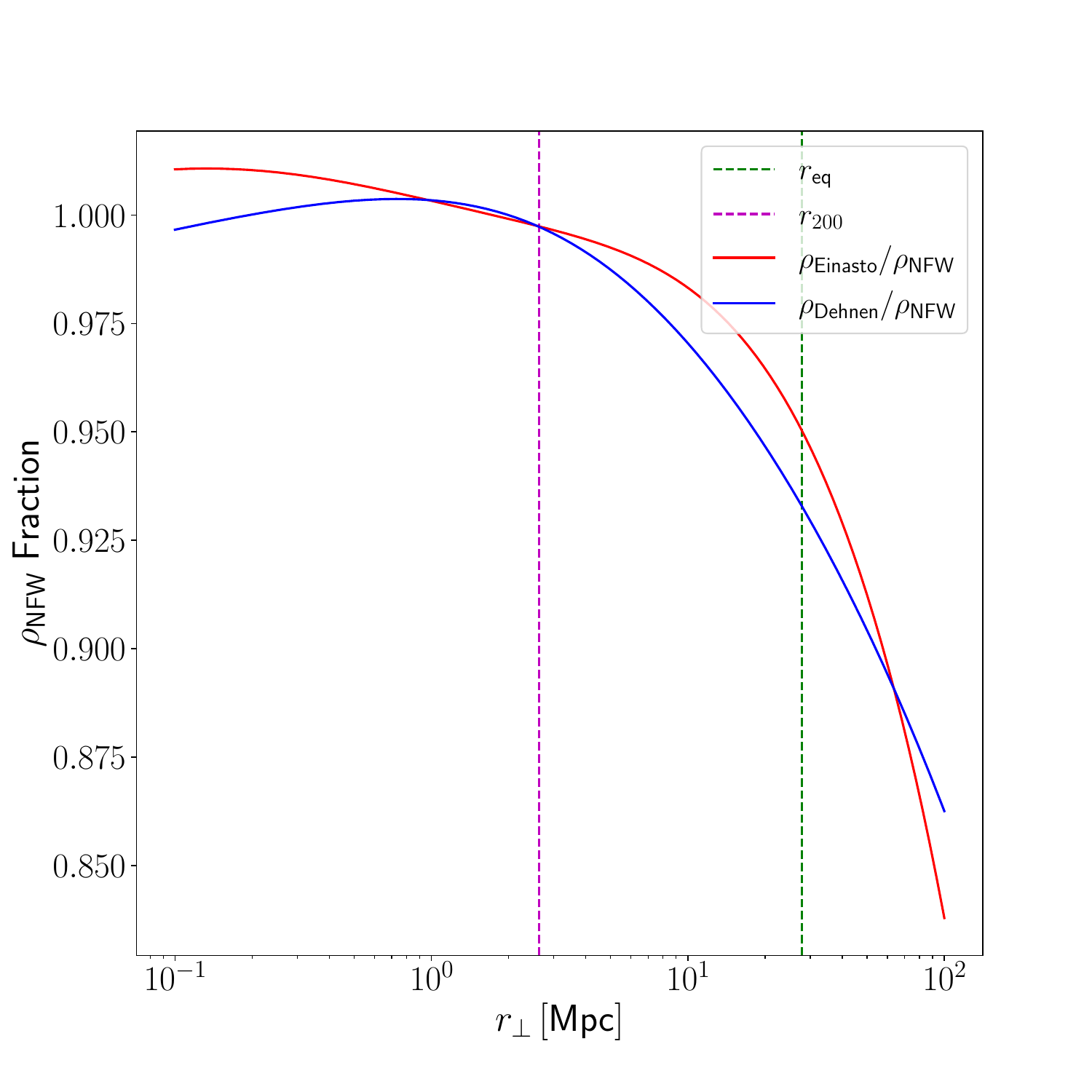}
\caption{\label{fig:density_profiles} The fraction of the Einasto or Dehnen density with respect to the NFW functional form for our best parameters in Table \ref{table:mass_estimates_this_work}. While there is close agreement inside $r_{200}$, the difference in the outer radii reflects the fact that both the Einasto and Dehnen are steeper. This leads to a smaller expected mass from the escape velocity profile.
} 


\end{figure}

Using the best-fit parameters, we plot the fractional difference in the density profiles in Figure \ref{fig:density_profiles}. We see percent level agreement out to $r_{200}$. Beyond that radius, we also see the Einasto and Dehnen densities become much smaller compared to the NFW, \arnote{expected from \citet{miller+2016}.} The key points of this figure are that\arnote{:} (a) the dynamically inferred density profiles are nearly identical within $r_{200}$ for the three functional forms we examine and (b) that the integral to $r_\text{{eq}}$ of the density in Equation \ref{eq:int_phi_poisson} will sum to a larger value in the NFW compared to the integrals for the Einasto and Dehnen forms. This is reflected in the estimated $M_{200}$ values in Table \ref{table:mass_estimates_this_work}, although the differences are not statistically significant. This analysis provides a promising new path forward in determining the true relative steepness of the densities in the outer regions of galaxy clusters, which is vital in studies of the splashback radius \citep{Contigiani19}.

\section{Velocity Dispersion-Based Mass}
\label{sec:Velocity Dispersion Analysis}


\arnote{One of the main techniques used to infer dynamical masses is through the velocity dispersion.} \citet{gomez+2012} estimated the velocity dispersion of AS1063 to be $\sigma_v=1840^{+230}_{-150}$ km/s within $\sim 730$ kpc. Due to the fact that they observed only a \arnote{small} fraction of the virial radius, they suggested a correction to this measurement out to $r_{200}$ based on an isothermal King model. Their corrected dispersion was reported as $\sigma_v=1660^{+230}_{-150}$ \arnote{km/s}, where they used the \citet{evrard+2008} mass vs. $\sigma$ relation for dark matter to estimate a mass \arnote{significantly higher than lensing and SZ estimates, which we report in Table \ref{table:Summaray_mass_estimates}}.

\arnote{More recently}, \citet{mercurio+2021} estimated $\sigma_v$ using the same data we use in this work, and found $\sigma_v=1380^{+26}_{-32}$ km/s for \arnote{the whole observational sample}. \arnote{However, this estimate cannot be easily translated into a virial mass given that the spectroscopic completeness varies significantly over the entire radial range, therefore making this dispersion difficult to interpret.}

\arnote{A more sophisticated dynamical approach was taken by \citet{sartoris+2020}, who performed a reconstruction of the mass profile through a Jean's likelihood regression analysis for the Jean's equation of dynamical equilibrium using the MAMPOSSt framework \citep{Mamon+2013}. From their mass profile, their virial mass is smaller than the \citet{gomez+2012} dispersion estimate, but both are systematically still larger than weak lensing and SZ measurements in Table \ref{table:Summaray_mass_estimates}. Moreover, the Jean's likelihood analysis implicitly assumes dynamical equilibrium \citep{Mamon+2013}, while AS1063 has been shown to have a notable presence of substructure \citep{gomez+2012,mercurio+2021}}. This motivates a further exploration of the dispersion-inferred mass for AS1063.
\arnote{
\subsection{The Dispersion of AS1063}
\label{subsec:The dispersion of A1063}
}

\arnote{We start by calculating the dispersion of AS1063, in this case following the same approach we used to estimate the phase-space edge profile by incorporating interloper stochasticity.  Previous dispersion estimates \citep{gomez+2012,mercurio+2021} of the system used a fixed bin/GAP pair to identify interlopers. The procedure is analogous to~\S\ref{subsec:mass_estimation}, except in this case we include data within $0.2\,r_{200}$ for the dispersion estimate, although no interlopers are identified within $0.2\,r_{200}$.}

We provide an estimate for the dispersion using standard bi-weighted estimator \citep{beers+1990}, of \arnote{$\sigma_v=1477^{+31}_{-30}$} km/s. The uncertainty on the dispersion is quantified using a bootstrapping technique \citep{beers+1990}. In our case, we use a Monte Carlo approach. First, we draw a new subset of galaxy velocities from the original data using normal distributions defined by the mean velocity of each galaxy and a fixed uncertainty of $150 \text{km/s}$ \citep{mercurio+2021}. We then bootstrap resample the data to create a new distribution of the velocities which accounts for the redshift measurement uncertainties.

We note that the bi-weight estimator requires a tuning parameter, {\it c} which defines the weights on the outliers. The value of the tuning parameter corresponds to the multiple of the median absolute deviation of the data. We examined the dispersion as a function of {\it c} and find that it depends strongly on {\it c} below {\it c} = 25. A high tuning constant is a better choice when few outliers are expected. Given that we employ the shifting-gapper technique to remove interlopers prior to calculating the dispersion, we argue that the tuning parameter should be defined with a value where it does not drive the actual value of the measured dispersion. Therefore, we choose {\it c} = 25. Values as low as {\it c} = 15 or as high as {\it c} = 30 change the measured dispersion by less than $0.3\%$ from its measurement for {\it c} = 25. However, the change in the dispersion is tens times that level (towards higher dispersion) going from {\it c} = 15 to {\it c} = 7.  \arnote{When using the default tuning constant used in Astropy \citep{astropy:2013,astropy:2018,astropy+2022}, of {\it c} = 9 , not accounting for proper weighting of outliers, we recover the same dispersion for the whole sample ($\sigma_v=1380^{+26}_{-32}$ km/s) as \citet{mercurio+2021}.}

\subsection{From $\sigma_{{\rm v, los}}$ to $\sigma_{{\rm v}}$}

\begin{figure*}
\includegraphics[width=.5\textwidth]{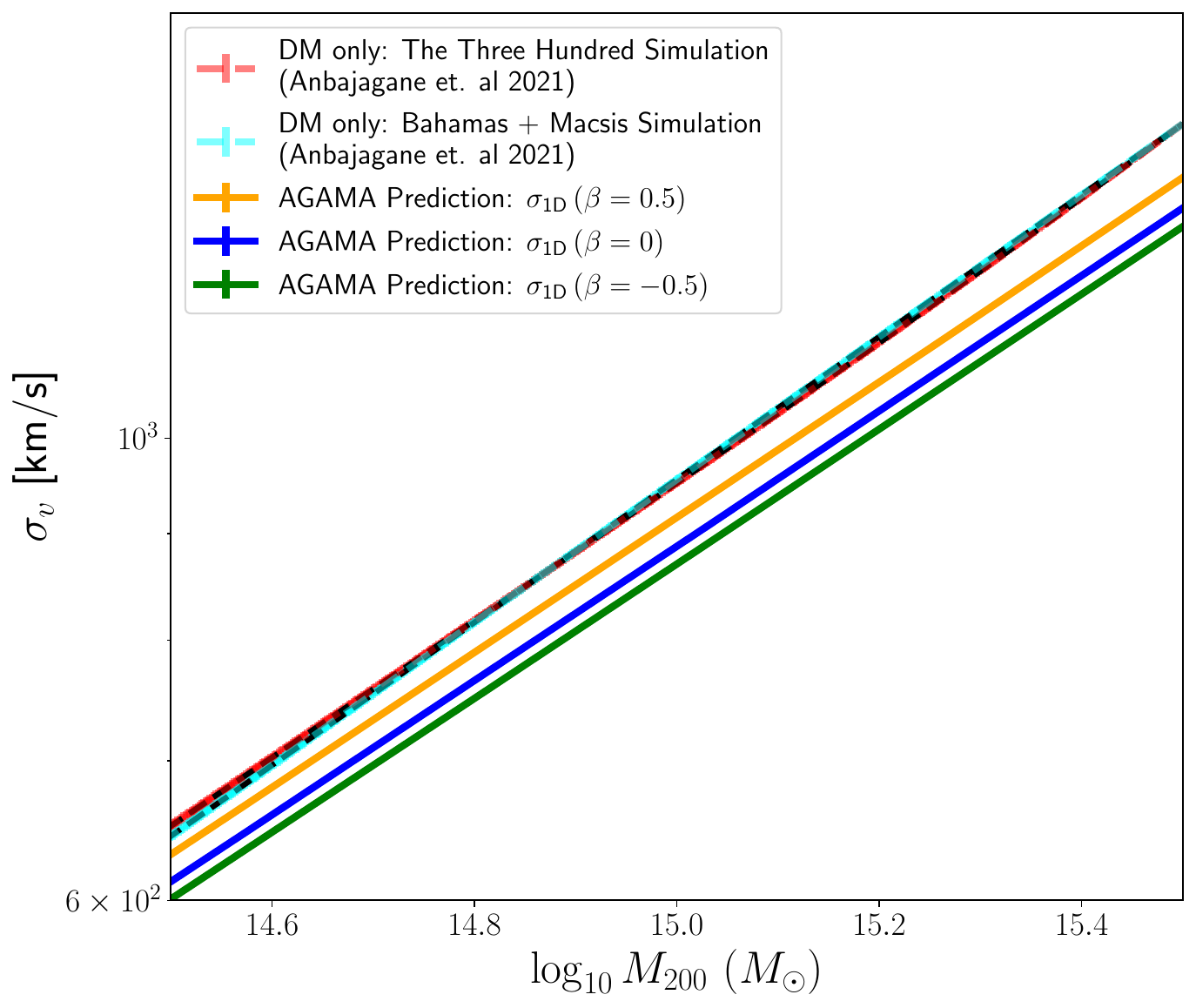}
\includegraphics[width=.5\textwidth]{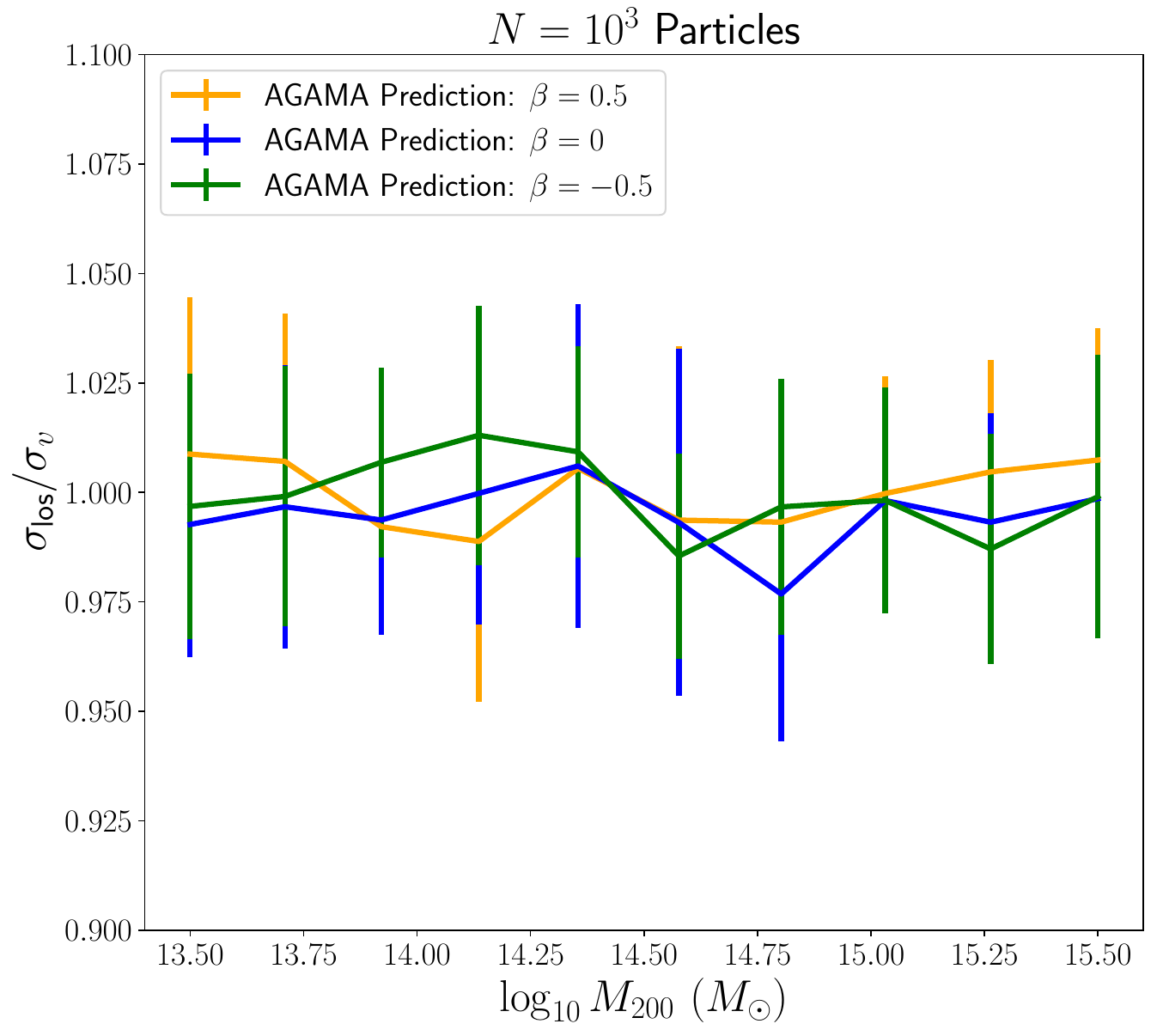}
\caption{\label{fig:v_los_show} $\textit{Left panel}$: Predicted virial scaling relation in AGAMA as a function of velocity anisotropy for $\beta=0.5$ (orange line), $\beta=0$ (blue line), and $\beta=-0.5$ (green line), where the small errorbars come from variation in phase-spaces for the $N=10^5$ particles. We find $\sim 5\%$ variation between the scalings between $\beta=0.5$ and $\beta=-0.5$.
 $\textit{Right panel}$: Ratio between the line-of-sight velocity dispersion and 1D velocity dispersion for $N=10^3$ particles. We find that when sampling along the line-of-sight, anisotropy and cluster mass have little effect, with $\sigma_{v,{\rm los}}$ recovering $\sigma_v$ in all cases (at higher N, all $\sigma_{v,{\rm los}}/\sigma_v$ converge to exactly 1). At this choice of $N$, we expect variation in this ratio of $4\%$, corresponding to an uncertainty in the line-of-sight dispersion of $\sim 60$ km/s for AS1063.
}
\end{figure*}

\begin{figure}
\includegraphics[width=.47\textwidth]{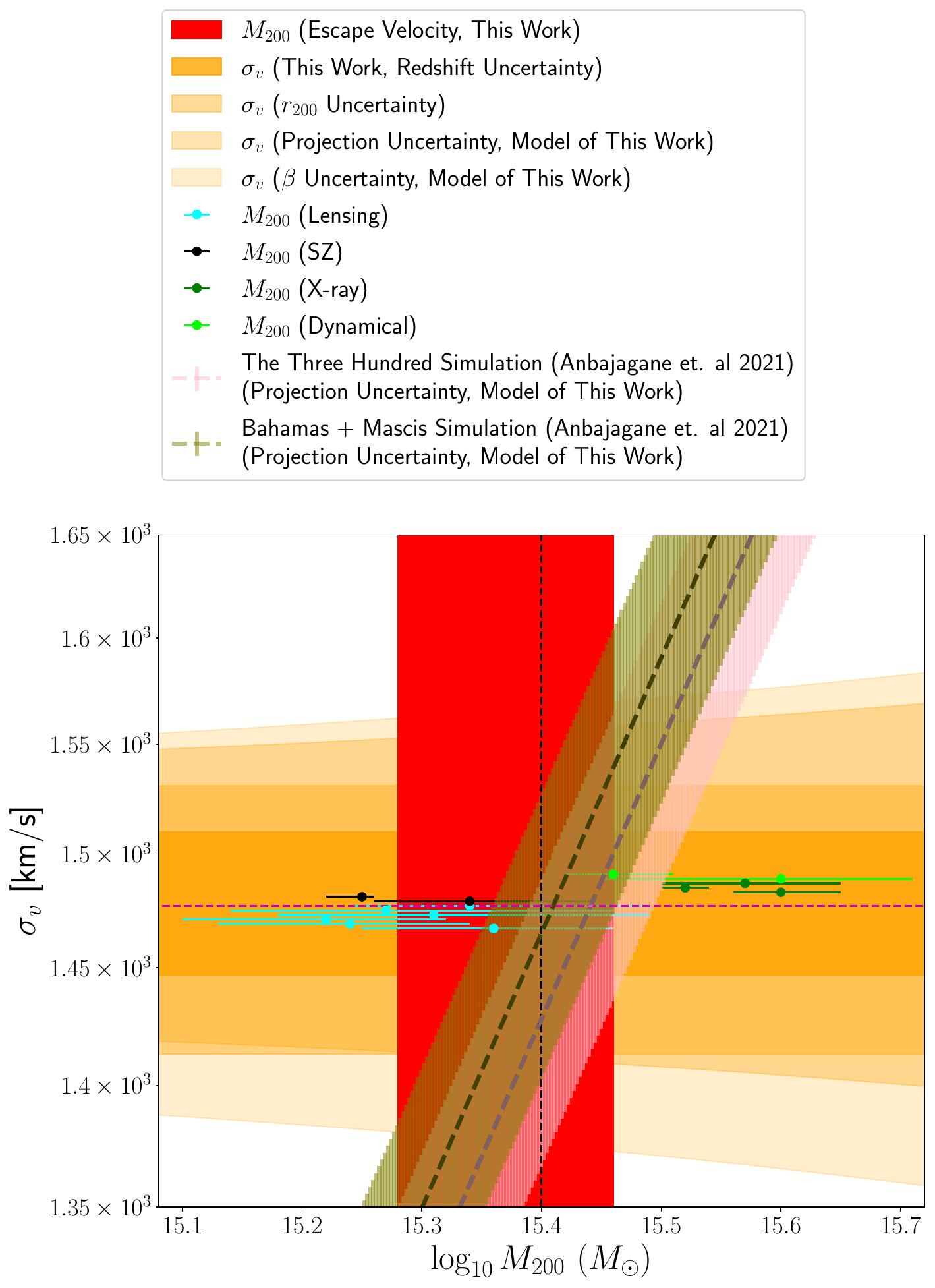}
\caption{\label{fig:v_los_data} Comparison of previous mass estimates in the literature from Table \ref{table:Summaray_mass_estimates} for lensing (cyan points), SZ (black points), X-ray (green points), and dynamical masses (lime green points). \arnote{We note these do not each have measured dispersions, but are plotted with values close to our $\sigma_v$ for convenience.} These are shown compared to the mass constraint of this work using the escape velocity, shown by the red shaded region. We also compare these to theoretical predictions from simulations from \citet{Anbajagane+2022}. We lastly show our velocity dispersion, measured from (1) The uncertainty determined through the $150$ km/s redshift error \citep{mercurio+2021} determined through bootstrap simulations (2) The uncertainty in using a line-of-sight dispersion given our sampling of $N=828$ tracers to measure the dispersion, using the projection model of this work (3) The uncertainty due to $\beta$, assuming $0<\beta<0.5$, yielding an additional contribution based on the model from this work.
}
\end{figure}

\arnote{Next, we perform dynamical modeling of cluster phase-spaces to investigate their more general properties. Our approach for dynamical modeling relies on the Action-based Galaxy Modeling Architecture (AGAMA) framework \citep{Vasiliev2019} (see~\S\ref{subsec:binning}). AGAMA can be contrasted with the Jean's inversion performed in \citet{sartoris+2020}, as AGAMA uses a predefined set of characteristics such as the system's mass, the density profile, the velocity anisotropy, etc. to infer the phase-space properties, rather than trying to infer these properties from the phase-space directly. In our case, we vary these parameters to see how they affect predictions of the velocity dispersion.} 

For a given M$_{200}$, we generate mock phase-space tracers at $z=0$. We use a variable number of tracers, from a few hundred to $N=10^{5}$. We specify a Dehnen density profile which is similar in shape to AS1063. We then define a \arnote{log-linear} grid of masses between $10^{14}$ and $10^{15.5}$ M$_{\odot}$, and also a coarse grid of constant velocity anisotropies between $\beta=-0.5$ and $\beta=0.5$ \citep{Lemze+2012, Stark+2019}. For the generated mock phase-space, the 1D velocity dispersion \arnote{$\arnote(\sigma_{\text{3D}}/\sqrt{3}$)} is inferred for the particles within $r_{200}$.


We first compare the AGAMA 1D dispersions to the simulation prediction for dark matter in halos from the Bahamas + Mascis Simulation \citep{McCarthy+2016,Barnes+2017} and The Three Hundred Project \citep{Cui+2019}, using the publicly available fitting functions from \citet{Anbajagane+2022}. \arnote{Three major differences between AGAMA and the N-body simulations are the non-linear growth patterns, realistic sub-structures, and a cosmological background, which all exist in the simulation halos.} We trade this for the ability to systematically control the velocity anisotropy of the mock clusters in AGAMA. This would be difficult in the simulations, where there is a finite (and often small) number of available halos which have galaxies with the right mixtures of orbits and dynamical populations. 

In the left panel of Figure \ref{fig:v_los_show}, we show the \arnote{predicted $\sigma_v$ vs. $M_{200}$ scalings} of the mock AGAMA phase-spaces when using $N=10^{5}$ tracers,  which is similar to the number of dark matter particles in the simulations.  Overall, we note three interesting findings: (1) the simulation particles have higher dispersions at fixed mass compared to the AGAMA tracers; (2) the slopes between the simulations and the AGAMA tracers differ; and (3) we find an unexpected dependence for the 1D dispersion by the anisotropy. Explaining these results is beyond the scope of this work. For now, we simply see how $\beta$ affects the 1D velocity dispersion at a given $M_{200}$. 

In the right-hand panel of Figure \ref{fig:v_los_show} we show how well we can recover the 1D dispersion ($\sigma_v$) using only projected velocities for cluster with the same number of tracers (\arnote{828} between $0.0 \le r_{\perp}/r_{200} \le 1$) as our dataset for  AS1063.   We find that our line-of-sight dispersion is unbiased as both a function of mass and anisotropy. 



\arnote{\subsection{$\sigma_v$ and spectroscopic completeness}}

\arnote{As noted earlier, \citet{mercurio+2021} report a non-uniform sampling completeness function corresponding roughly to: $100\%$ from $0\,r_{200}$ to $0.1\,r_{200}$, $80\%$ from $0.1\,r_{200}$ to $0.5\,r_{200}$, and $70\%$ from $0.5\,r_{200}$ to $r_{200}$. A benefit of AGAMA is that we can create mock phase-spaces using this sample spectroscopic completeness function in order to access the magnitude of its affect on the dispersion.}

\arnote{The phase-spaces we construct are analogous to AS1063, where we use $N=828$ particles within the virial radius, and estimate the line-of-sight dispersion using 100 line-of-sight draws from the specified completeness function. We find the statistical uncertainty from the phase-space sampling (see~\S\ref{subsec:sigma and its uncertainties}) dominates over the effect of non-uniform completeness, where we find a difference of $4\pm34$ km/s compared to a full completeness function for a $\beta=0.5$ cluster, where differences in $\beta$ are negligible compared to the statistical uncertainty. Due to the fact that the statistical uncertainty dominates over the completeness uncertainty, we neglect the effect of non-uniform sampling on the dispersion analysis.}

\centerline{}
\subsection{$\sigma_v$ and its uncertainties}
\label{subsec:sigma and its uncertainties}

We made an initial estimate of the dispersion in~\S\ref{sec:membership}, where we found \arnote{$\sigma_v=1477^{+31}_{-30}$} km/s for galaxies within the virial radius.  As noted before, we choose a tuning constant \citep{beers+1990} of $c=25$, corresponding to $25\times$ the median absolute deviation from the median residual in the data. This high tuning constant results in a stable measurement of the dispersion (see~\S\ref{subsec:The dispersion of A1063}). The errors are calculated as before using bootstrap resampling and accounting for galaxy redshift errors \arnote{of 150 km/s}. As shown in the last section, we expect $\sigma_{v, {\rm los}}$ to be an unbiased measure of the 1D velocity dispersion, $\sigma_v$. 

We consider two additional effects which introduce uncertainties into our determination of $\sigma_v$: sampling, and radial dependence.
We calibrate a model for how sampling affects the projection. This has been previously studied in simulations by other authors (\cite{Gifford2013, Serra2011,Caldwell+2016}. 
\arnote{We find that the statistical scatter as a function of phase-space sampling follows the same simple power-law obtained by \citet{Caldwell+2016} who used simple Gaussians as the 1D representatives of the phase-space data.} We measure a finite sampling uncertainty of $^{+43}_{-43}$ km/s from given our sample size of \arnote{$N=828$} within r$_{200}$. The size of this error is reflected in the right panel of Figure \ref{fig:v_los_show}.


Since $\sigma_v$ is only measured within $r_{200}$, the uncertainty in the inferred virial radius presents an additional source of error. When constructing the mock line-of-sight dispersion profile in AGAMA, we find no considerable differences in $\sigma_v(<r)$ in the radial profile between $0.2\,r_{200}$ and $r_{200}$. However, in the data we find differences of $^{+21}_{-34}$ km/s between the inner and outer 1$\sigma$ ranges on r$_{200}$. We include these as a systematic uncertainty.



Our final dispersion uncertainty on the inferred 1D $\sigma_v$ from the observed $\sigma_{v,{\rm los}}$ includes (1) the redshift measurement uncertainty; (2) the systematic error from the uncertainty in $r_{200}$;
(3) the finite sampling uncertainty; and (4) the uncertainty in theory from anisotropy from Figure \ref{fig:v_los_show} (left). The final dispersion we obtain is then \arnote{$\sigma_{v} =1477^{+87}_{-99}$} km/s.



In Figure \ref{fig:v_los_data}, we plot the escape mass \arnote{using the analytic suppression model}, inferred by the NFW density profile (vertical red bands) against the inferred 1D velocity dispersion (horizontal orange bands). \arnote{If we used the Millennium suppression model, we infer more suppression corresponding to lower masses by$~\sim0.1$ dex}. The colored squares are mass values from the literature (\arnote{see} Table \ref{table:Summaray_mass_estimates}). The literature data are placed at the position of the central value of $\sigma_v$ and offset for clarity.
The diagonal bands are expectations for galaxies with stellar masses greater than $5\times 10^{9}M_{\odot}$ from two hydrodynamic simulations \citep{Anbajagane+2022}.


When compared to masses inferred from weak (and strong) lensing, the SZ effect, and the X-ray temperature, we find that $\sigma_v$ is \arnote{consistent with the expectations from the cosmological hydrodynamic simulations so long as the escape velocity mass of this work is used (note again the masses are not directly associated with a $\sigma_v$, but are simply plotted for convenience in a band similar to our own), with most lensing estimates being within $1\sigma$. The Jean's estimate is also in good agreement with the theoretical prediction, while only one of the SZ masses appears to be. The X-ray masses are slightly above the theoretical expectations, and adds to the evidence that the system is not in hydrostatic equilibrium.} 



\section{Discussion}
\label{sec:discussion}
Our goal in this work is to infer an accurate and precise mass for AS1063 using the escape velocity. As a baseline, we focus our comparisons on non-dynamical lensing inferred masses. There are other studies comparing masses (or mass profiles) inferred from the ``caustic'' technique to weak lensing \citep{Diaferio2005, Geller13, Rines2013, Hoekstra2015, Herbonnet2020}. \arnote{However, this is the first attempt at measuring the mass of a galaxy cluster using the escape edge as a tool and as derived in \citet{nandra+2012}, \citet{Behroozi+2013}, \citet{miller+2016}, and \citet{halenka+2020}. This technique is fundamentally different than the ``caustic'' technique \citep{diaferio+1997, Gifford_2013} for three reasons: (1) cosmology is a required component in our analysis; (2) the suppression of the edge of the data compared to dark matter simulations is assumed to be caused from statistical sampling and not from the velocity anisotropy \citep{halenka+2020}; (3) the velocity dispersion is not used to calibrate the edge.}

In our analyses, we have been careful to study the uncertainties on the escape mass, which include not only measurement error on the galaxy redshifts, but also edge error from the interloper removal process and uncertainty in the radial suppression. We have paid similar attention to the error in the inferred 1D velocity dispersion.

In terms of precision, we have shown that this new technique provides similar or better mass constraints compared to current weak lensing data. Surprisingly, the \arnote{covariance} between M$_{200}$ and concentration has both a different shape and also a lesser effect on the mass than in weak lensing measurements (see Figure \ref{fig:phase_space}). \arnote{We suggest that this lack of sensitivity in the constraint of the shape of the potential is from the fact that the escape profile is the square root of the potential profile, which is naturally flatter than the steep density profile.} This opens up a new possibility to place more precise constraints on the mass-concentration relation \arnote{through joint shear/v$_{\text{esc}}$ posterior constraints on mass and shape.} This idea might also enable better theoretical predictions on the M$_{200}$-c relation, since the potential and density profiles of dark matter particles in simulations are independent measures of the underlying shape of the Poisson-pair profile.

\arnote{In order to infer the mass from the edge, we use a statistical suppression model based on the work of \citet{halenka+2020}. The value of this suppression can be measured using an analytic model for the phase-space or N-body simulations. In terms of the radially averaged value, the predictions are similar ($\frac{1}{Z_v} = 0.78$ and $\frac{1}{Z_v} = 0.84$ for the analytic and N-body models respectively). This is an additional systematic that needs further evaluation. Our reported masses in Table \ref{table:mass_estimates_this_work} and all of the figures use the AGAMA calibration. Had we instead used the Millennium-based suppression, our final mass for AS1063 would be lower by about 0.1dex and close aligned to the weak-lensing and SZ data in Figure \ref{fig:v_los_data}.}

Prior work meanwhile (see~\S\ref{subsec:suppression}) had explained the edge suppression through velocity anisotropy via $g(\beta)$\arnote{$^{-2}$} in Equation \ref{eq:DG1997}. Thus, we have that $g(\beta)$\arnote{$^{-2} \equiv$} $ \frac{1}{Z_v} = 0.78$\arnote{$^{+0.05}_{-0.04}$}.  
However, a nominal value of $\beta = 0.25$ would suppress the observed edge \arnote{by 0.55, which is much more than what we observe.} Pure isotropy would suppress it a little less to \arnote{0.58}. A fully rotating cluster (i.e. with no radial motion of its members) would suppress the edge to 70\% \arnote{of its 3D value, which is closer to what we observer, but \arnote{is} unphysical}. The difference is exacerbated with the Millennium-based suppression value. From this we conclude that the explanation for suppression of the observed edge due to velocity anisotropy is inconsistent with both the data and the theory. 



For Abell S1063, we have shown that it is possible to its dynamical mass estimate for a non-relaxed system in-line with weak lensing and SZ mass estimates, \arnote{as well as with theoretical predictions from the Three-Hundred Simulation and Bahamas + Mascis hydrodynamic cosmological simulations \citep{McCarthy+2016,Barnes+2017,Cui+2019}}

\arnote{On a final note, even though the weak lensing and escape masses are consistent with each other, they may still be individually biased. Consider the weak lensing mass estimate of \citet{Umetsu_2016}. Besides the statistical uncertainty on the shear measurement, the total cluster mass uncertainties incorporate systematic errors due to the mass-sheet uncertainty, uncorrelated large-scale structure, halo asphericity, the presence of correlated halos, model-dependent systematics in the strong lensing, as well as systematics from their radial binning scheme. While an exhaustive list, there is the possibility that unknown systematics remain, such as internal cluster substructure. \citet{gruen2013} identified a second density peak within 1 Mpc of the center and regressed against a mass model which contained two NFW halos. However, their combined two halo mass is statistically consistent with their single NFW model fit. }

\arnote{For the escape mass, the mass uncertainty incorporates statistical uncertainties on the galaxy redshifts, systematic uncertainties from the shifting-gapper parameters, intrinsic scatter in the suppression function, and the effect of realistic phase-space sub-structure on the suppression function. Like lensing, it is possible that there remain unknown systematics, such as internal substructure. However, \citet{Behroozi+2013} used simulations to show that the escape velocity has negligible correlation with the presence of substructure. Therefore, other than radial binning, there are no clear systematics shared in common between weak lensing and the escape techniques. To fully quantify the agreement between an escape-based dynamical mass and a shear-based non-dynamical mass, we need to take the next steps to increase the sample size and we need clusters with quality data for both the lensing and the spectroscopy.}

\section{Acknowledgments}
This material is based upon work supported by the National Science Foundation under grant No. 1812739.
\footnotesize{
\bibliographystyle{natbib}
\bibliography{master}
}

\input{appendix.tex}

\end{document}

%% file: appendix.tex
\appendix
\renewcommand{\thefigure}{A\arabic{figure}}
\label{sec:appendix}

\setcounter{figure}{0}
\section{Alternate Density Profiles}
In this Appendix, we present the fits for the additional density profiles, where~\S\ref{subsec:mass_estimation} only presented the fits to the NFW. As outlined in~\S\ref{subsec:mass_estimation}, we use the $Z_v(r)$ from Figure \ref{fig:Zv_use} to account for radial suppression, and a degree-two polynomial smoothing function for the phase-space edge to account for over and under-fitting.

Here, we consider the Einasto \citep{einasto+1969,montenegro+2012} and Dehnen \citep{dehen+1993} proifles. The NFW is most distinct from the other two profiles in that it falls off less quickly in the outskirts \citep{navarro+1997}. The potential for the Einasto can be obtained by solving the Poisson equation, and is given by \citep{montenegro+2012}:
\begin{align}
\label{eq:Einasto}
    \Psi(r)&=\frac{-GM_{\text{tot}}}{r}\left[1-\frac{\Gamma(3n,(\frac{r}{h})^{1/n})}{\Gamma(3n)}+\frac{r}{h}\frac{\Gamma(2n,(\frac{r}{h}))^{1/n}}{\Gamma(3n)}\right]
\end{align}
where $\Gamma(\alpha,x)$ is the incomplete gamma function, given by $\Gamma(\alpha,x)=\int_{x}^{\infty}t^{\alpha-1}e^{-t}dt$. The total mass, $M_{\text{tot}}$, meanwhile is given by $M_{\text{tot}}=4\pi\rho_0 h^3 n \Gamma(3n)$, and here  $\rho_0$ is a normalization factor, $h$ is the scale radius, and $n$ is the Einasto index. Meanwhile, the potential for the Dehnen is  \citep{dehen+1993}:
\begin{align}
\label{eq:Dehnen}
    \Psi(r)&=\frac{GM_{\text{tot}}}{r_s}\frac{-1}{2-\gamma}\left[1-(\frac{r}{r+r_s})^{2-\gamma}\right], n\neq2
\end{align}
\begin{align}
    \Psi(r)&=\frac{GM_{\text{tot}}}{r_s}\log(\frac{r}{r+r_s}), n=2
\end{align}
where the total mass $M_{\text{tot}}$ is again a normalization factor, $r_s$ is the scale radius, and $\gamma$ is the Dehnen index.

\begin{figure*}
\includegraphics[width=.5\textwidth]{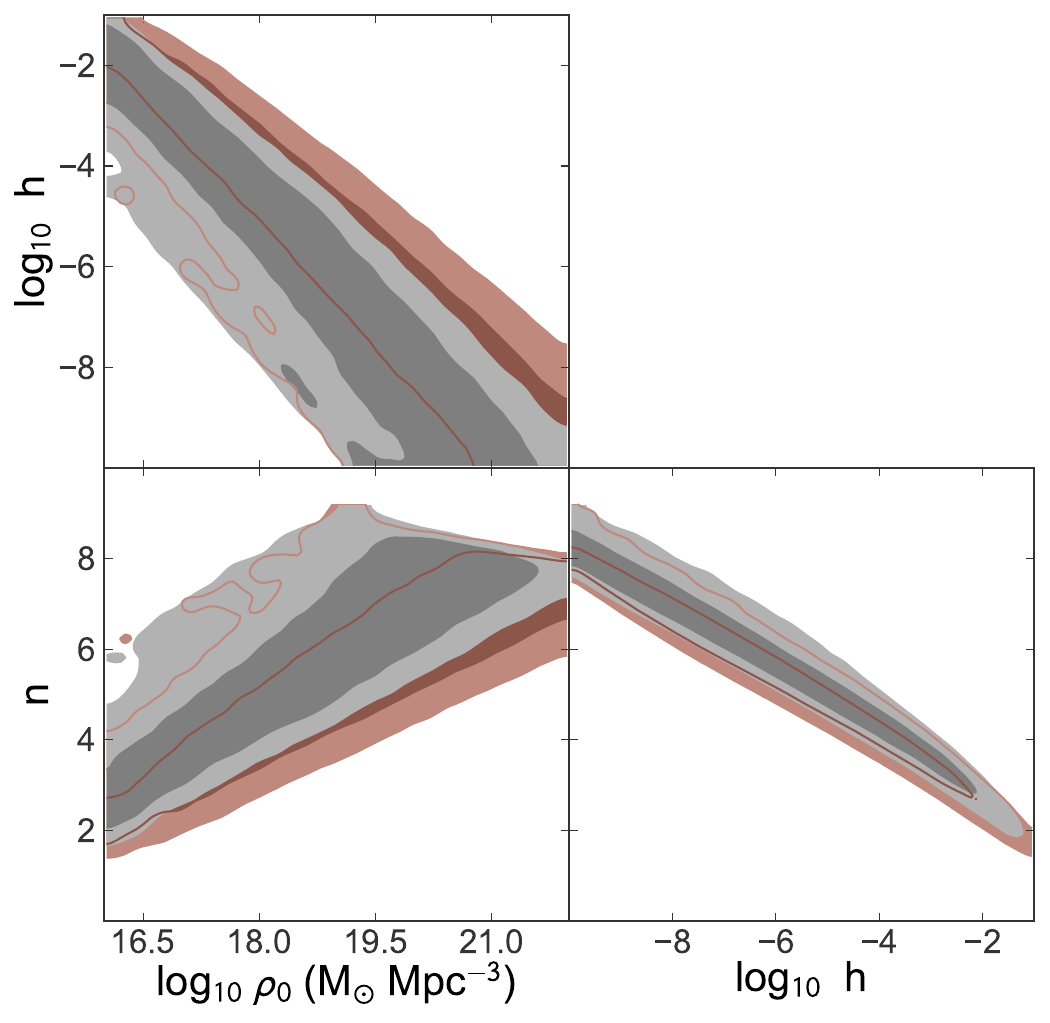}
\includegraphics[width=.5\textwidth]{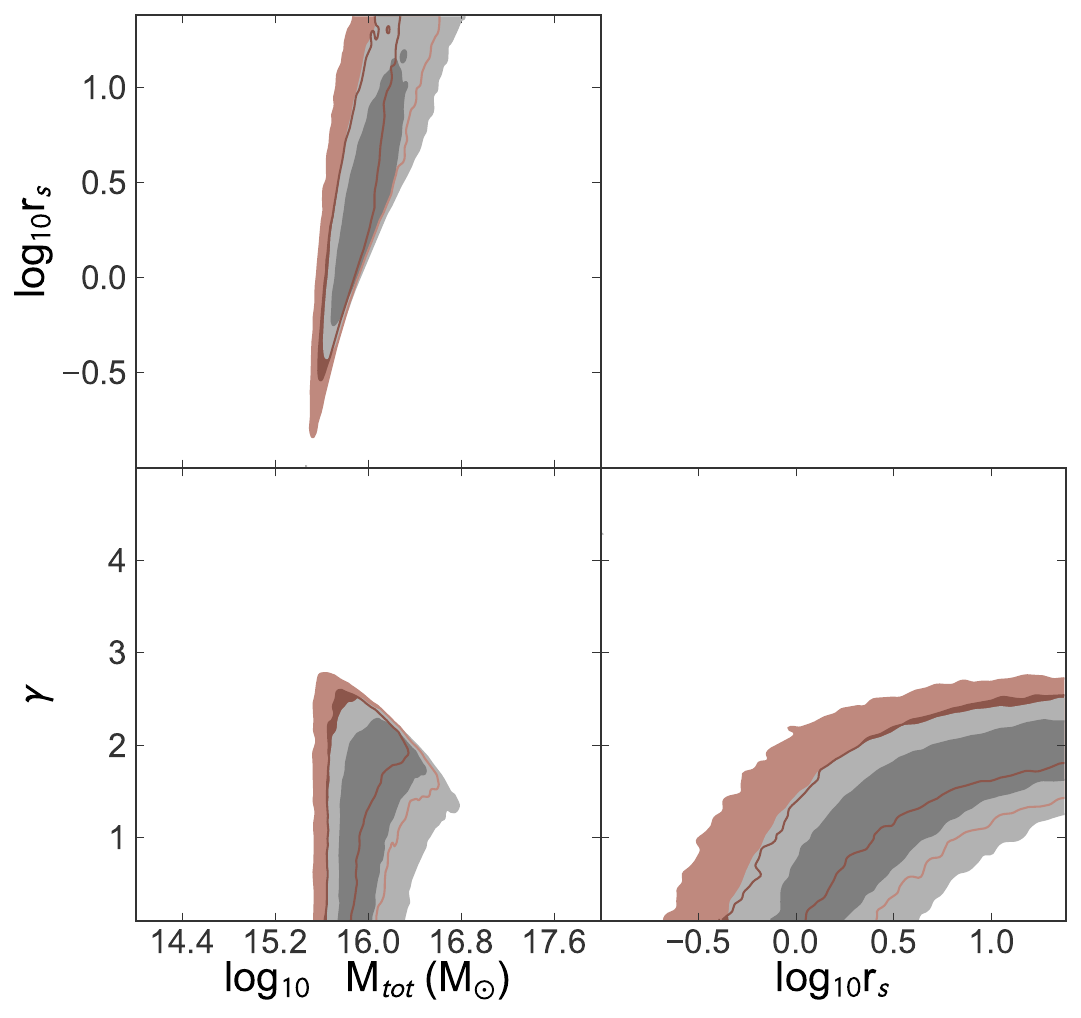}\\
\includegraphics[width=.5\textwidth]{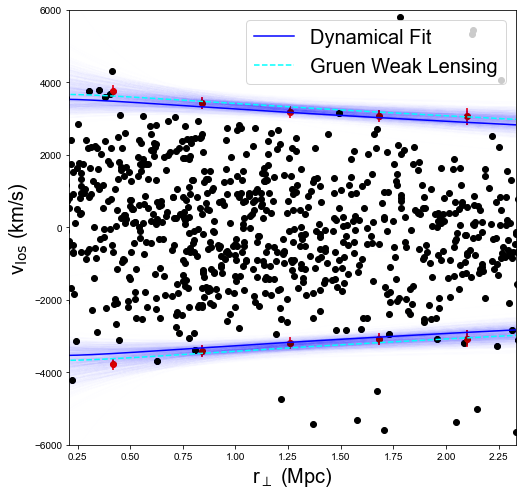}
\includegraphics[width=.5\textwidth]{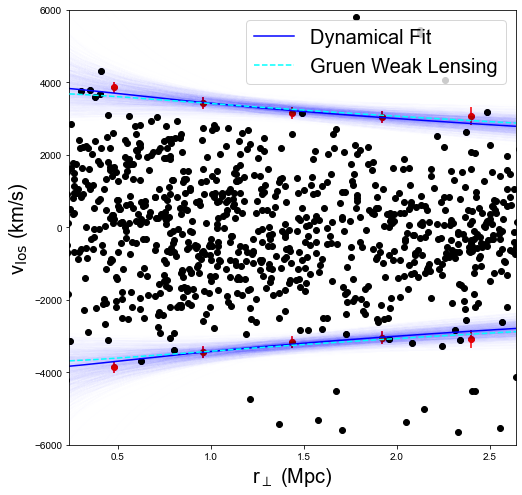}

\caption{\label{fig:Posterior_figures} \textit{Top panel}: 2D posterior estimates  (68\% and 95\%) of the Einasto (left) and Dehnen (right) profile parameters. The covariance of the distribution suggests that only a narrow range of the parameters values yield physical mass constraints. However, the ranges are both wider than in the case of the NFW, seen in the top panel of Figure \ref{fig:phase_space}. \textit{Bottom row}: The escape profiles estimated from drawing from the posterior distributions. The \citet{gruen2013} weak lensing estimate (cyan lines) is in-line with our mass/concentration predictions for the two density profiles, as for the NFW.}

\end{figure*}

The top panel of Figure \ref{fig:Posterior_figures} shows the density profile posteriors for the Einasto (left) and Dehnen (right) profiles. In each case, priors for the free parameters are given in Table \ref{table:Summaray_priors}. We use the GTC package \citep{Bocquet+2016} to generate the histograms. The posteriors are highly degenerate, yet yield relatively tightly constrained masses within $\sim0.1$ dex. They are similar to the top left panel of Figure \ref{fig:phase_space}, in that only a slim range of the density profile's parameters yields a physical escape profile, just like the NFW. However in the case of the Einasto and Dehnen profiles, there is more degeneracy than the case of the NFW. The bottom panel of Figure \ref{fig:Posterior_figures} is analogous to the bottom panel of Figure \ref{fig:phase_space}, except using the additional density profiles. Like the NFW, there is clear agreement between the dynamical fits and the prediction from \citet{gruen2013}, and suggests that there is no clear distinction between any of the profiles, highlighted as well in Table \ref{table:mass_estimates_this_work}.

\begin{table}[t]
  \caption{The priors on the function parameters used in the potential profiles \arnote{(Equations \ref{eq:nfw_psi}, \ref{eq:Einasto}, and \ref{eq:Dehnen} for the NFW, Einasto, and Dehnen profiles respectively). The parameters are drawn from a uniform distribution within the specified ranges.}}
  \centering
  \begin{tabular}{lcr}
 \hline\hline
 NFW: \\ 
 \hline\hline
 $r_s$ & $0.1<r_s/\hat{R}_{200} <10 $ \\ 
 $\rho_s$ & $12<\log_{10}(\rho_s)<16$ \\
  \hline\hline
 Einasto: \\ 
 \hline\hline
 $\rho_0$ & $16<$log$_{10}(\rho_0)<22$ \\
 $h$ & $-10<$log$_{10}$(h)$<-1$\\
 $n$ & $1<n<10$\\
  \hline\hline
  Dehnen: \\ 
 \hline\hline
 $\gamma$ & $0.1<\gamma<5$\\
 $r_s$ & $0.1<r_s/\hat{R}_{200} <10 $ \\ 
 $M_{\text{tot}}$ & $14<\log(M_{\text{tot}})<18 $
  \end{tabular}
  
    \label{table:Summaray_priors}
\end{table}

